\documentclass[12pt,preprint]{aastex}

\usepackage{color}
\usepackage{amsmath}
\usepackage{fullpage}
\usepackage{subfigure}

\shorttitle{Modeling of the internal structure for surface shedding}
\shortauthors{Hirabayashi et al.}

\begin{document}


\title{Internal Structure of Asteroids Having Surface Shedding due to Rotational Instability}


\author{Masatoshi Hirabayashi\altaffilmark{1}}
\email{masatoshi.hirabayashi@colorado.edu}

\author{Diego Paul S\'anchez\altaffilmark{2}}

\and

\author{Daniel J. Scheeres\altaffilmark{3}}
\affil{Aerospace Engineering Sciences, 429 UCB, University of Colorado, Boulder, CO 80309-0429 United States}


\altaffiltext{1}{Research Associate, Colorado Center for Astrodynamics Research, 
Aerospace Engineering Sciences, University of Colorado Boulder}

\altaffiltext{2}{Senior Research Associate, Colorado Center for Astrodynamics Research, 
Aerospace Engineering Sciences, University of Colorado Boulder}

\altaffiltext{3}{Richard Seebass Chair, Professor, Colorado Center for Astrodynamics Research, 
Aerospace Engineering Sciences, University of Colorado Boulder}



\begin{abstract}
Surface shedding of an asteroid is a failure mode where {\it surface materials} fly off due to strong centrifugal forces beyond the critical spin period, while the internal structure does not deform significantly. This paper proposes a possible structure of an asteroid interior that leads to such surface shedding due to rapid rotation rates. A rubble pile asteroid is modeled as a spheroid composed of a surface shell and a concentric internal core, the entire assembly called the test body. The test body is assumed to be uniformly rotating around a constant rotation axis. We also assume that while the bulk density and the friction angle are constant, the cohesion of the surface shell is different from that of the internal core. First, developing an analytical model based on limit analysis, we provide the upper and lower bounds for the actual surface shedding condition. Second, we use a Soft-Sphere Discrete Element Method (SSDEM) to study dynamical deformation of the test body due to a quasi-static spin-up. In this paper we show the consistency of both approaches. Additionally, the SSDEM simulations show that the initial failure always occurs locally and not globally.  In addition, as the core becomes larger, the size of lofted components becomes smaller. These results imply that if there is a strong enough core in a progenitor body, surface shedding is the most likely failure mode.  
\end{abstract}





\section{Introduction}
The Yarkovsky-O'Keefe-Radzievskii-Paddack effect, so-called the YORP effect, play a crucial role in providing asymmetric asteroids with rotational torque due to the solar radiation pressure \citep{Rubincam2000}. This implies that asteroids with several hundred meters in diameter may be able to gradually spin up to their spin limits \citep{Rubincam2000}. Observational surveys have shown that there is a spin period threshold of $\sim 2.3$ hr for asteroids larger than a few hundred meters in size \citep{Pravec2007}. Recent observations detected spin-up/down of asteroids due to this effect \citep{Lowry2007, Taylor2007, Kaasalainen2007, Durech2008}. These reports imply that spin-up processes could cause asteroid structures and shapes to evolve over their life time.

Some recently observed disruption events of active asteroids are considered to have resulted from rotational instability; however there are other proposed mechanisms for these events as well. \cite{Jewitt2010} and \cite{Snodgrass2010} both independently reported the disruption event of asteroid P/2010 A2. \cite{Jewitt2010} concluded that it was likely that the formation of the debris tail resulted from rotational instability, while \cite{Snodgrass2010} noted that a collisional impact is a likely event driving such events. \cite{Jewitt2014B} described the recent breakup event of P/2013 R3, suggesting that this event came from rotational instability. \cite{Hsieh2004,Hsieh2010} reported that the dust tail of asteroid 133P/Elst-Pizarro, spinning with a spin period of 3.471 hr, may come from seasonal activity and rotational instability. \cite{Sheppard2015} also observed a thin dust tail of asteroid (62412) 2000 SY178 whose spin period is 3.33 hr from March 28 to May 2 in 2014. Although the activities of these bodies could be driven by sublimation of water ice that might be buried in the bodies, the slow particles in their dust tails also imply that rotational instability may also play a role in mass ejection from the surface. \cite{Jewitt2013, Jewitt2015_P5} conducted observations of P/2013 P5 (PANSTARRS) with the Hubble Space Telescope to investigate its episodic dust ejection, implying that this event might result from rotational instability. \cite{Hainaut2014} and \cite{Hainaut2014ACM} proposed an alternative scenario of this event in which the dust tail might be generated by the soft contact in a binary system gently rubbing each other.

Over the last decade, there have been many important analytical \citep{Holsapple2001, Holsapple2004, Holsapple2007, Holsapple2010} and numerical \citep{Walsh2008, Walsh2012, Sanchez2012} investigations of failure conditions of asteroids. Of specific relevance for this study, it has been found that for a homogeneous strength distribution in an asteroid, the internal structure is more sensitive to failure than the surface region \citep{Hirabayashi2015_DA}. Another active research area is an understanding of mass ejection from a fast rotating body due to centrifugal forces. A popular explanation for the origin of mass ejection is that surface materials flow down to the equator, and if the rotation is fast enough, they fly away. \cite{Guibout2003} investigated the surface stability due to the shape and spin period of a uniformly rotating, self-gravitating ellipsoid and found that based on the body configuration, the surface materials may accumulate at different ends. Using a Hard-Sphere Discrete Element Method (HSDEM), \cite{Walsh2008,Walsh2012} showed the formation of a binary system due to mass ejection from a cohesionless spheroid. By considering a constant angle of repose, \cite{Minton2008} and \cite{Harris2009} numerically obtained equilibrium shapes of a uniformly rotating symmetric body and obtained a body shape similar to 1999 KW4 Alpha. \cite{Scheeres2015} analytically investigated granular surface flows on rotating asteroids and their equilibrium shapes, demonstrating the formation of unique surface slopes of oblate shapes such as 1999 KW4 Alpha and 2008 EV5. Also, \cite{Hirabayashi2014} compared surface mass ejection and structural failure due to rapid rotation to determine a possible failure mode of (216) Kleopatra. 

However, the mechanism behind surfaces mass ejection is still poorly understood, and thus we will try to shed some light on it in this paper. Developing a two-bulk-density-layer model, \cite{Hirabayashi2014B} showed that a high-density core makes the whole body structurally strong. Assuming a homogenous structure, \cite{Hirabayashi2015_DA} conducted plastic finite element analysis to explain that the failure mode of asteroid (29075) 1950 DA is plastic deformation of the internal region. These studies imply that in order for the surface region of an asteroid to fail first at an elevated spin rate, its interior must be stronger than a surface layer. Based on these studies, we propose one way to create surface shedding from a rubble pile asteroid. Here, we introduce a simplified model to discuss it. Our rubble pile asteroid is modeled as a self-gravitating spheroid composed of two cohesive strength layers, a spherical shell and an internal core. The bulk density and the friction angles are assumed to be constant to simplify our discussion. We note that cohesion may depend on porosity \citep{Lambe1969}, but we will leave consideration of cohesion and porosity for a future study. 

In the following discussion, we will use the term ``surface shedding." So, before we do, we will try to give it a proper definition. \cite{Hirabayashi2014} introduced this term to explain the mode that particles just resting on the surface are shed due to strong centrifugal forces. However, this definition is still ambiguous. First, if particles on the surface interact mechanically with other particles by cohesion, it is difficult to determine the lofting condition by simple balance between gravitational and centrifugal forces. Second, if cohesion is uniformly distributed over the whole volume, the internal structure deforms before the force balance point, i.e., the dynamical equilibrium point, reaches the surface \citep{Holsapple2010, Hirabayashi2015_DA}. To define this mode clearly, we describe ``surface shedding" as a mode in which {\it while the internal structure is below the yield condition, the surface region fails structurally and particles in this region are shed due to strong centrifugal forces.} We hypothesize that this mode may play a role in the disruption events of 133P/Elst-Pizarro \citep{Hsieh2004,Hsieh2010}, P/2013 P5 \citep{Jewitt2013, Jewitt2015_P5} and 62412 \citep{Sheppard2015}. 

\section{Analytical Modeling of Mechanical Failure}
\subsection{Internal Core Model}
\begin{figure}[hb]
  \centering
  \includegraphics[width=6in]{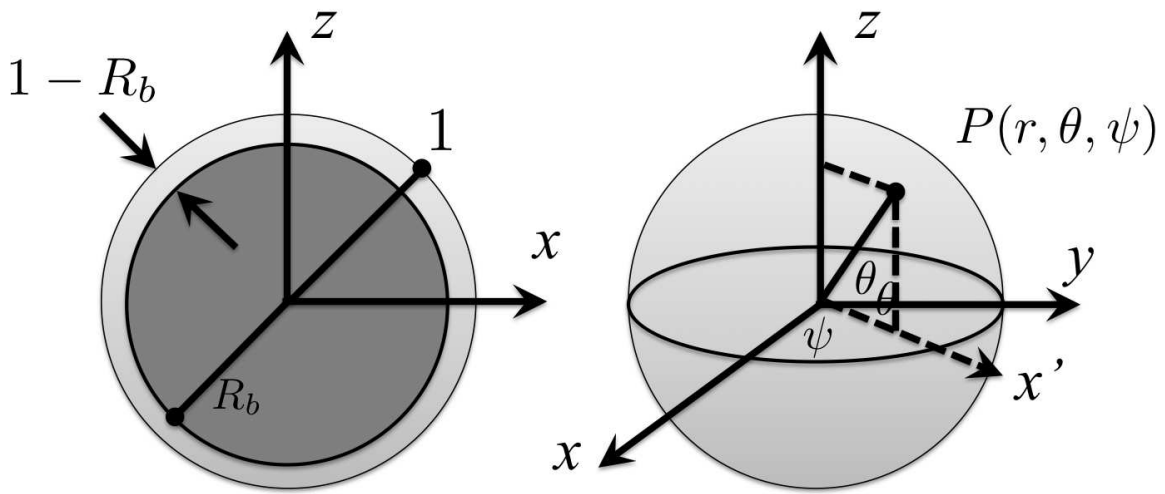}
  \caption{Sphere model with the internal core. The sphere is assumed to be spinning constantly along the $z$ axis. The normalized radii of the sphere and the internal core are given as $1$ and $R_b$, respectively.}
  \label{Fig:sphereModel}
\end{figure}

We suppose that the test body considered here is a uniformly rotating, self-gravitating spheroid that has a concentric, spherical core under the surface shell, later known as the internal core (Figure \ref{Fig:sphereModel}). It is assumed that these layers have different cohesion, while their bulk density and friction angle are the same over the whole volume. Surface shedding is induced not by the failure of the internal core but by that of the surface shell. If the test body is spinning fast enough, surface particles are shed from the test body; otherwise, they just move over the surface. 

The bulk density, the total radius and the gravitational constant are denoted as $\rho$, $R$ and $G$, respectively. We normalize lengths, body forces, spin rates and stress tensors by $R$, $\pi \rho G R$, $\sqrt{\pi \rho G}$ and $\pi \rho^2 G R^2$, respectively. With this normalization, the sphere radius is denoted as $1$ and the radius of the internal core is defined as $R_b$, so the thickness of the surface shell is given as $1-R_b$. The right plot in Figure \ref{Fig:sphereModel} shows the spherical coordinate system $(r,\theta,\phi)$, but we also use the cartesian coordinate system $(x, y, z)$ as well. The spin axis is assumed to be constant along the $z$ axis. We consider that at a given spin rate, if the surface shell already fails structurally, the cohesion between the particles there is broken and they can move freely. For this case, the particle is lofted from the surface due to the centrifugal force at a spin rate of $\sqrt{4/3} \sim 1.15$, while it energetically escapes to infinity at that of $\sqrt{8/3} \sim 1.63$. 

\subsection{Failure Conditions of the Two Layers}
\label{Sec:LimitAnalysis}
In the analytical model developed here, the behavior of materials is assumed to be elastic-perfectly plastic and to follow an associated flow rule. The yield condition of a material is assumed to be characterized by the Mohr-Coulomb yield criterion, a pressure-shear dependent yield criterion, which is given as 
\begin{eqnarray}
(\sigma_1 - \sigma_3) \sec \phi + (\sigma_1 + \sigma_3) \tan \phi \le 2 Y. \label{Eq:MC}
\end{eqnarray}
where $\sigma_i \: (i=1,2,3)$ is the principal stress component, $\phi$ is a friction angle and $Y$ is cohesion. Since we consider a sphere, the stress state is always $\sigma_1 = \sigma_2 \ge \sigma_3$. 

The yield conditions of the surface shell and the internal core are calculated by using an elastic stress solution. Solving the elastic equations, including the equilibrium equations, the constitutive equations (Hooke's law), the strain-displacement equations and proper boundary conditions (zero-traction condition), we obtain the solution as \citep{Dobrovolskis1982}
\begin{eqnarray}
\sigma_{xx} &=& k_1 (1 - r^2) - k_{13} x^2 - k_{14} z^2, \label{Eq:sigxx} \\
\sigma_{yy} &=& k_5 (1 - r^2) - k_{13} y^2 - k_{15} z^2, \\
\sigma_{zz} &=& k_9 (1 - r^2) - k_{14} x^2 - k_{15} y^2, \label{Eq:sigzz} \\
\sigma_{xy} &=& k_{13} x y, \label{Eq:sigxy} \\
\sigma_{xz} &=& k_{14} x z, \\
\sigma_{yz} &=& k_{15} y z, \label{Eq:sigyz}
\end{eqnarray}
where $r^2  = x^2 + y^2 + z^2$ and the coefficients, $k$s, are defined as follows:
\begin{eqnarray}
k_1 &=& \frac{2 b_x (-12 - \nu + 5 \nu^2) + b_z (3- 6\nu - 5 \nu^2)}{10 (-1 + \nu) (7 + 5 \nu)}, \label{Eq:k1} \\
k_9 &=& \frac{2 b_x (3 - 6 \nu - 5 \nu^2) + b_z (-27 + 4 \nu + 15 \nu^2)}{10 (-1 + \nu) (7 + 5 \nu)}, \label{Eq:k9} \\
k_{13} &=& - \frac{2 b_x (-4 + 3 \nu + 5 \nu^2) + b_z (1 + 3 \nu)}{5 (-1 + \nu) (7 + 5 \nu)}, \label{Eq:k13} \\
k_{14} &=& - \frac{b_x (-3 + 6 \nu + 5 \nu^2) + b_z (-4 + 3 \nu + 5 \nu^2)}{5 (-1 + \nu) (7 + 5 \nu)}. \label{Eq:k14}
\end{eqnarray}  
Also, $k_5 = k_1$ and $k_{14} = k_{15}$. $\nu$ is Poisson's ratio. The coefficients of the body force components along the $x$, $y$ and $z$ axes, denoted as $b_x$, $b_y$ and $b_z$, respectively, are given as 
\begin{eqnarray}
b_x = b_y = \omega^2 - \frac{4}{3}, \:\:\:\:\:\: b_z =  - \frac{4}{3}. 
\end{eqnarray}

To determine lower and upper bound conditions for the actual failure, we apply the upper and lower bound theorems in limit analysis \citep{Chen1988,Chen1990}. At a lower bound condition, the target volume should not fail structurally. At an upper bound condition, on the other hand, the volume should fail. Thus, the actual failure condition should always be between these conditions. 

The lower bound theorem describes that when (1) an equilibrium stress solution is found and (2) there is a small element whose stress reaches its yield condition, the body may not widely fail. This means that the yield condition of a small element does not guarantee further propagation of failure regions. Therefore, we say that this condition is below the actual failure condition. Practically, this condition is given as the yield condition of the stress at the most sensitive element. Because the test body is spherical, such an element always appears on the equatorial plane. 

For the lower bound condition of the surface shell, the boundary between the surface shell and the internal core becomes the most sensitive to failure. Considering $y=z=0$, we obtain the principal stress components at this location as 
\begin{eqnarray}
\sigma_{e1} &=& k_1 (1-r^2), \\
\sigma_{e2} &=& k_1 (1-r^2) - k_{13} r^2, \\
\sigma_{e3} &=& k_9 (1-r^2) - k_{13} r^2.
\end{eqnarray}
It is not always satisfied that $\sigma_{e1} > \sigma_{e2} > \sigma_{e3}$. Thus, to have the correct order, we calculate $\sigma_{es1} = \max (\sigma_{e1}, \sigma_{e2}, \sigma_{e3})$ and $\sigma_{es3} = \min (\sigma_{e1}, \sigma_{e2}, \sigma_{e3})$. On the other hand, for the lower bound condition of the internal core, since the center is always the most sensitive location, we obtain the principal stress components as 
\begin{eqnarray}
\sigma_{ec1} = k_1, \:\:\: \sigma_{ec2} = k_1, \:\:\: \sigma_{ec3} = k_9.
\end{eqnarray}
The lower bound condition provides the highest approximated value of cohesion for the actual failure condition, later known as the highest cohesion. Based on Equation (\ref{Eq:MC}), this is expressed as 
\begin{eqnarray}
Y_{ek} = \frac{1}{2} (\bar \sigma_{ek1} - \bar \sigma_{ek3}) \sec \phi + \frac{1}{2} (\bar \sigma_{ek1} + \bar \sigma_{ek3}) \tan \phi, \label{Eq:Yek}
\end{eqnarray}
where $k$ is either $s$ or $c$. $k = s$ indicates cohesion of the surface shell, while $k=c$ provides that of the internal core.

For the upper bound theorem, on the other hand, using the theory by \cite{Chen1988} and by \cite{Chen1990}, \cite{Holsapple2008A} developed a practical technique. His theorem guarantees that on the assumption that the velocity field is linear, for a given volume, the yield condition of the averaged stress over it is equal to an upper bound condition for the actual failure. The averaged principal stress over a given volume $V_k$ is given as
\begin{eqnarray}
\bar \sigma_{jk} = \frac{1}{V_k} \int_{V_k} \sigma_{j} dV_k. \label{Eq:avgStressFormula}
\end{eqnarray}
Here, we average the stresses over the surface shell and the internal core. Using Equation (\ref{Eq:avgStressFormula}), we calculate the averaged stress components of the surface shell as
\begin{eqnarray}
\bar \sigma_{s1} &=& \bar \sigma_{s2} = - k_1 \frac{R_b^3 - R_b^5}{1 - R_b^3} + \frac{b_x}{5} \frac{1 - R_b^5}{1 - R_b^3}, \label{Eq:sig1s} \\
\bar \sigma_{s3} &=& - k_9 \frac{R_b^3 - R_b^5}{1 - R_b^3} + \frac{b_z}{5} \frac{1 - R_b^5}{1 - R_b^3}, \label{Eq:sig2s}
\end{eqnarray}
and those of the internal core as 
\begin{eqnarray}
\bar \sigma_{c1} &=& \bar \sigma_{c2} = k_1 \left(1 - R_b^2 \right) + \frac{b_x R_b^2}{5}, \label{Eq:sig1in} \\
\bar \sigma_{c3} &=& k_9 \left(1 - R_b^2 \right) + \frac{b_z R_b^2}{5}. \label{Eq:sig2in}
\end{eqnarray}
The details of these equations are in Appendix \ref{App:AvgStress}. To verify these stress components, we show that they recover the averaged stresses over the whole volume obtained by \cite{Holsapple2007}. If $R_b = 0$, the averaged stress over the surface shell times $\pi \rho^2 G R^2$ is identical to Equation (4.12) with $a = b = c = R$ in \cite{Holsapple2007}. If $R_b = 1$, Equation (4.12) is recovered again since this case is reduced to aggregates with a homogeneous interior. This consistency allows us to compare the upper bound conditions of the surface shell and the internal core. The yield conditions of these stress components provide the lowest estimated value of the actual cohesion, later known as the lowest cohesion. Similar to Equation (\ref{Eq:Yek}), this is described as
\begin{eqnarray}
Y_{k} = \frac{1}{2} (\bar \sigma_{k1} - \bar \sigma_{k3}) \sec \phi + \frac{1}{2} (\bar \sigma_{k1} + \bar \sigma_{k3}) \tan \phi. \label{Eq:Yk}
\end{eqnarray}

In Section \ref{Sec:FailureCondition}, we discuss the upper and lower bound conditions for the surface shell and the upper bound condition for the internal core. For the internal core, since the surface shedding mode is described by a core that does not reach the yield, the lower bound condition for the internal core does not provide constraints on it. Therefore, we do not consider this condition in the present problem. 

\section{Numerical Modeling of Surface Granular Flow}
\label{Sec:numerical}
The simulation program that is used for this research applies a SSDEM to simulate a self-gravitating granular aggregate \citep{cundall1971, cundall1992, sanchez2011}. The particles, modeled as spheres that follow a predetermined size distribution, interact through a soft-repulsive potential when in contact.  This method considers that two particles are in contact when they overlap.  When this happens, normal and tangential contact forces are calculated \citep{herr1}. 

The calculation of the normal forces between colliding particles is modeled by a hertzian spring and a dashpot.  The elastic force is modeled as
\begin{equation}
{\vec{\bf f}}_e= k_n\xi^{3/2}{\bf\hat n},
\label{hook}
\end{equation}
the damping force as
\begin{equation}
{\vec{\bf f}}_d=-\gamma_n\dot\xi{\bf\hat n},
\end{equation}
and the cohesive force between the particles is calculated as
\begin{equation}
{\vec{\bf f}}_c=-2 \pi \frac{r_1^2 r_2^2}{r_1^2 + r_2^2} \sigma_{yy} \hat {\bf r}_{12}
\end{equation}
where $r_1$ and $r_2$ are the radii of the two particles in contact, $\sigma_{yy}$ is the tensile strength of this contact, which is given by a cohesive matrix formed by the (non simulated) interstitial regolith \citep{sanchez2014}, and $\hat{\bf{r}}_{12}$ is the branch vector between the centers of these two particles. Then, the total normal force is calculated as ${\vec{\bf f}}_n={\vec{\bf f}}_e+{\vec{\bf f}}_c+{\vec{\bf f}}_d$.  In these equations, $k_n$ is the elastic constant, $\xi$ is the overlap of the particles, $\gamma_n$ is the damping constant (related to the dashpot), $\dot\xi$ is the rate of deformation and ${\bf\hat n}$ is the normal vector on the surfaces of the colliding particles.  As the simulation only deals with spheres, $\hat{\bf r}_{12}$ has the direction as $\hat {\bf n}$. This dashpot models the energy dissipation that occurs during a real collision.

The tangential component of the contact force models surface friction statically and dynamically. This is calculated by placing a linear spring attached to both particles at the contact point at the beginning of the collision \citep{herr1,silbert} and by producing a restoring frictional force ${\vec{\bf f}}_{t}$. The magnitude of the elongation of this tangential spring is truncated in order to satisfy the local Coulomb yield criterion $|{\vec{\bf f}}_t|\leq\mu |{\vec{\bf f}}_n|$. 


Rolling friction \citep{ai-chen2011} has also been implemented in order to mimic the behavior of aggregates formed by non-spherical grains. That is, particles are subjected to a torque that opposes the relative rotation of any two particles in contact.  This torque, similar to surface-surface friction, is implemented as linearly dependent on the relative angular displacement of any two particles in contact and has a limiting value of:
\begin{equation}
M_r^m=\mu_r R_r |{\vec{\bf f}}_n| 
\end{equation}
where $\mu_r$ is the coefficient of rolling resistance, $R_r=r_1r_2/(r_1+r_2)$ is the rolling radius and $r_1$ and $r_2$ are the radii of the two particles in contact. This allows our simulations to reach friction angles of up to $\approx 35^{\rm o}$ as evaluated by the Druker-Prager yield criterion \citep{Sanchez2012}.  This value is typical of geological aggregates, though friction angles of $\sim$40$\rm^o$ are not rare. This implementation of rolling friction is similar to that of surface friction, but is instead related to the relative angular displacement. 

To obtain the initial configuration of the aggregates, we leave the particles (initially cohesionless and frictionless) to coalesce only under the influence of their mutual gravitational interactions.  Then, the particles are encapsulated inside a perfectly solid sphere that shrinks to a predetermined size.  This size has been pre-determined from previous simulations so that the particles are forced into a close-to-perfect spherical shape.  When the set-up procedure is finished, friction and cohesive forces are applied and the simulation is ready to start. Once the simulation starts, the aggregate is spun up in small, discrete increments of a normalized spin rate of $5.5\times 10^{-3}$ every 3000 s around the spin axis and the disruption process is observed.  This time  step showed to be long enough to allow the aggregates to safely reshape before the next spin-up event.  The aggregates rotate in the -z direction to facilitate the observation of the reshaping and disruption process through the graphical interface of the simulation code.

For this research, to observe the effect of an internal core in a self-gravitating aggregate, we have chosen to have cores whose radii are 0.5, 0.6, 0.7, 0.8 and 0.9. The particles forming the core are subjected to cohesive forces that are five times as strong as those felt by the particles of the shell. All the aggregates used in the simulations are formed by 3000 perfectly spherical particles with normalized radii between 0.025 - 0.035. The normalized parameters of the model are as follow: $k_n=6.1\times 10^7$, $k_t=2.1\times 10^6$, $\gamma_n=67$, $\mu=0.5$ and $\mu_r=0.8$.

\section{Analysis of Surface Shedding}
\subsection{Failure Condition of the Internal Structure}
\label{Sec:FailureCondition}
With the analytical model developed above, we now investigate the effect of the internal core on the surface failure and the surface shedding condition. The friction angle is fixed at 35$^\circ$, a mean value of typical friction angles of geological materials \citep{Lambe1969}. For a choice of Poisson's ratio, we account for the earlier studies by \cite{Holsapple2008A} and by \cite{Hirabayashi2014}. They confirmed that Poisson's ratio do not critically change the failure condition\footnote{More specifically, they compared two Poisson's ratios, 0.2 and 0.33, and confirmed that the critical spin conditions only changed within $2 \%$.}. Here, to take into account compressibility of a soil material, we will define Poisson's ratio as 0.25. With these fixed values, we derive the upper and lower bound conditions for the surface shell and the upper bound condition for the internal core. It is noted that since the surface shedding mode is characterized by an internal core being below the yield and a failing surface shell, the lower bound condition for the internal core does not give any constraint on cohesion. 

\begin{figure}[ht!]
	\begin{center}
		\subfigure[]{
         		\label{Fig:Surface}	
		\includegraphics[width=3.2in]{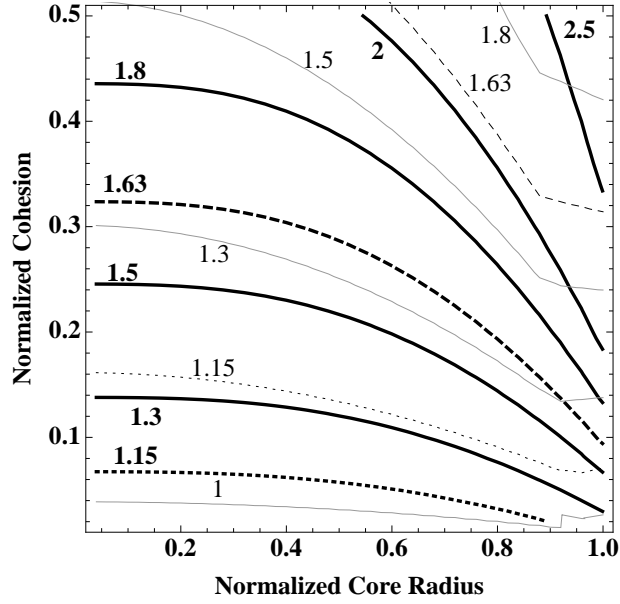}
          	} \\
		\subfigure[]{
         		\label{Fig:Core}	
		\includegraphics[width=3.2in]{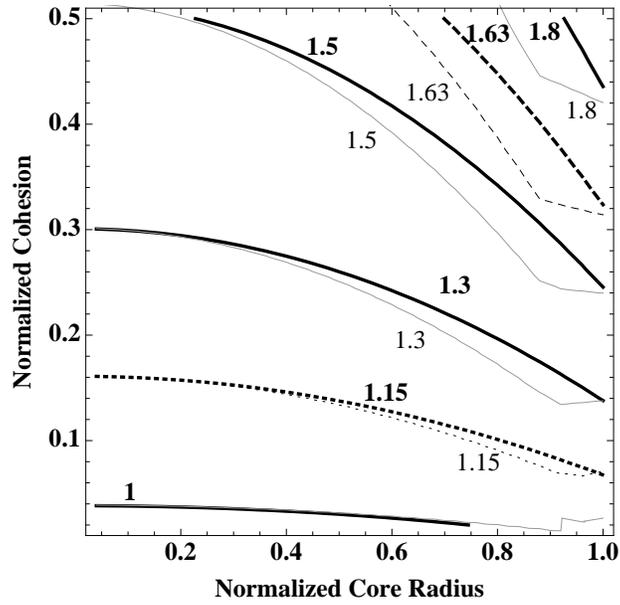}
          	}
	\caption{Normalized cohesion as a function of the normalized core radius. The numbers are the normalized spin states, which are constant on each line. Figure \ref{Fig:Surface} shows the highest (bold lines) and lowest (narrow lines) cohesion of the surface shell. Figure \ref{Fig:Core} describes the highest cohesion of the internal core (bold lines) and the lowest cohesion of the surface shell (narrow lines). The dotted lines indicate the spin rate at which a particle is just lofted, $\sim1.15$, while the dashed lines describe that at which a particle escapes to infinity, $\sim 1.63$.}
	\label{Fig:minPeriod}
	\end{center}
\end{figure}

Figure \ref{Fig:minPeriod} shows the upper and lower conditions for the normalized cohesion as a function of the normalized core radius. The spin rate is constant on each line. We have also plotted the 1.15 spin rate, at which a particle resting on the surface can be gently lofted, and the 1.63 spin rate, at which a particle escapes from the gravity of the test body. The former and latter spin rates are given by the dotted and dashed lines, respectively. Figure \ref{Fig:Surface} gives the failure condition of the surface shell. The bold lines define the lowest cohesion at a given spin rate, while the narrow lines provide the highest cohesion at that spin rate. The actual failure spin rate should be between these two lines. At a constant spin rate, as the core radius increases, the highest and lowest cohesion decrease. On the other hand, though the surface shell should fail structurally, the internal core should be below the yield (see Figure \ref{Fig:Core}). The bold lines describe the lowest cohesion of the internal core at a given spin rate, while the narrower lines shows the highest cohesion of the surface shell at that spin rate. These results show that at a constant spin rate, the lowest cohesion that the internal core needs to remain stable is higher than the cohesion that the surface shell requires to be able to fail. This means that if the cohesion of the internal core is not high enough, it will tend to fail before the surface.  In a homogeneous body, failure should start therefore at its center and propagate towards the surface. This scenario corresponds to the possible failure mode of asteroid 1950 DA \citep{Hirabayashi2015_DA}.

Figures \ref{Fig:Surface} and \ref{Fig:Core} imply that if surface shedding is observed, details of the process could provide information about their internal structure. Given a constant spin period, we could give constraints on cohesive strength. Since the actual cohesion should be enclosed by the lowest and highest cohesion curves, it should be possible to obtain the range of cohesion based on a possible range of the core radius. In general, the core radius is not well determined; however, it is still possible to give a constraint on it. For example, let us consider a test body that experiences surface shedding at a spin rate of 1.3. For the surface shell (Figure \ref{Fig:Surface}), taking the minimum value of the lowest cohesion (0.04 at a core radius of 1.0) and the maximum value of the highest cohesion (0.3 at that of 0), we roughly obtain the range of cohesion of the surface shell as 0.04 - 0.3. On the other hand, for the internal core (Figure \ref{Fig:Core}), we roughly give that of the internal core as $>0.14$. 

\subsection{Dynamics of Failure Mode}
This section investigates dynamics of failure modes by changing the size of the internal core. The initial conditions of the friction angle and the porosity are given based on experimental and observational evidences. The friction angle considered in the numerical simulation is 35$^\circ$, which is consistent with the analytical model. The porosity is calculated to be 34$\%$, similar to that of the lunar regolith observed by Apollo 15 mission \citep{Mitchell1972}. This is calculated by dividing the total mass of the particles into the volume obtained from semi-axes of a dynamically-equivalent equal-volume ellipsoid (DEEVE) with the same moments of inertia as the aggregate. 

\begin{figure}[ht!]
	\begin{center}
		\subfigure[]{
         		\label{Fig:core00a}	
		\includegraphics[width=3.1in]{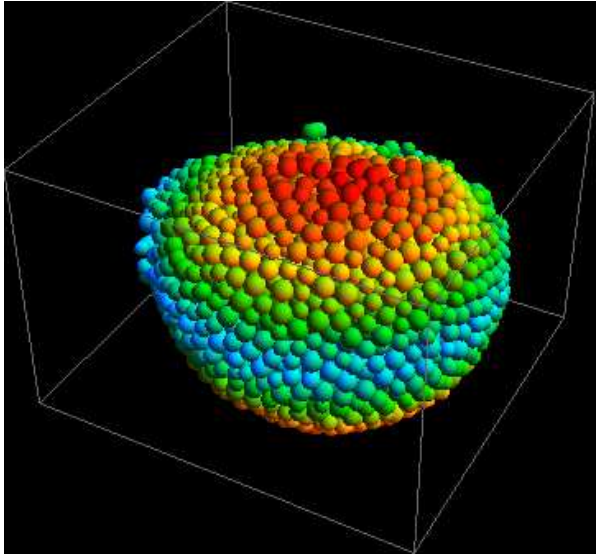}
          	}
		\subfigure[]{
         		\label{Fig:core00b}	
		\includegraphics[width=3in]{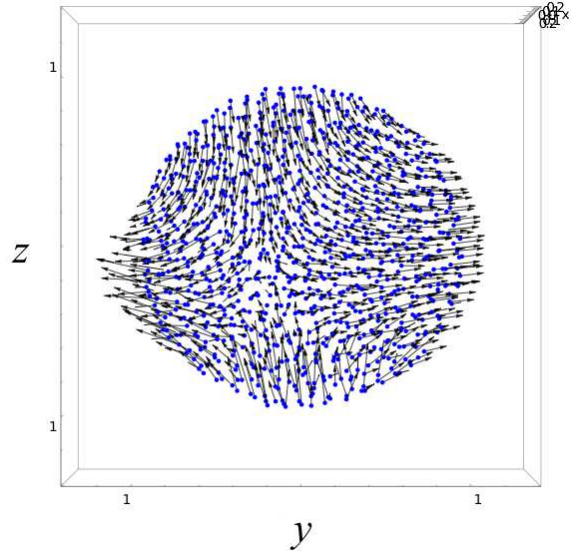}
          	} \\
		\subfigure[]{
         		\label{Fig:core00c}	
		\includegraphics[width=2.85in]{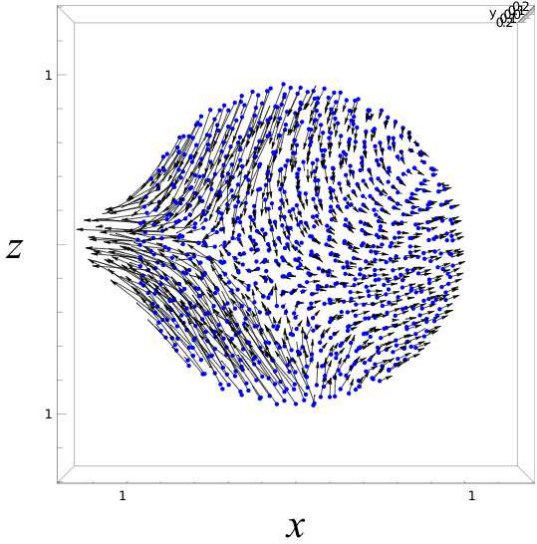}
          	}
		\subfigure[]{
         		\label{Fig:core00d}	
		\includegraphics[width=3in]{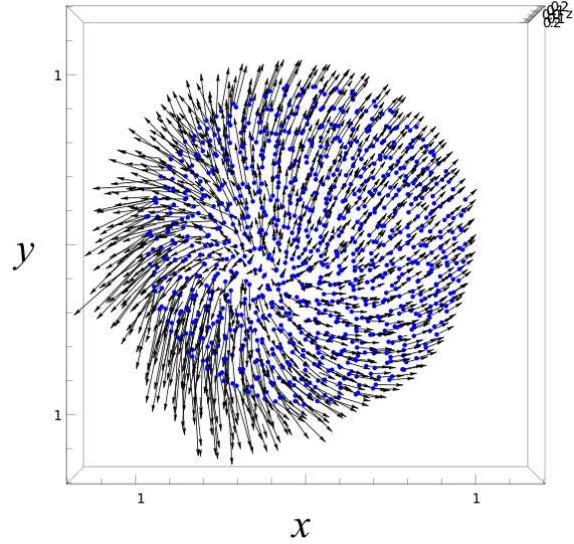}
          	} \\	
	\caption{Deformation of the test body with a normalized core radius of 0. Thus, the normalized cohesion of the whole body is 0.1. The normalized critical spin rate is 1.13. Figure \ref{Fig:core00a} shows the 3-dimensional view. The color shows the latitude of the test body. Figures \ref{Fig:core00b} through \ref{Fig:core00d} show the displacement vectors of the particles contained in thin slices perpendicular to the x, y and z axes as seen from +x, +y and +z respectively.  In these pictures, the blue dots are the original locations of the particles.}
	\label{Fig:core00}
	\end{center}
\end{figure}

The normalized cohesion of the internal core and the surface shell is 0.5 and 0.1, respectively\footnote{Since cohesion is given in units of pressure, i.e., stress components, we normalize it by $\pi \rho^2 G R^2$.}. To obtain these values, we conduct the following numerical settings; given the tensile strength between the particles of the shell, $\sigma_{ts}$, the tensile strength between the particles in the core is described as $\sigma_{tc}=5\sigma_{ts}$ and that between the particles in the interface is obtained as $\sigma_{ti}=1.5\sigma_{ts}$. These values of cohesion were chosen as previous simulations of homogeneous aggregates showed their critical spin rates and disruption modes to be very different; a 0.1 cohesion homogeneous aggregate fails at 1.13 (Figure \ref{Fig:core00}) and a 0.5 cohesion aggregate breaks at 1.44 (Figure \ref{Fig:ZeroCore_Ultimate}). These figures are explained later. Also, as seen in Figure \ref{Fig:minPeriod}, from the analytical model, known are the spin rates at which homogeneous aggregates would deform and disrupt, and so the ratio of cohesive forces warrants that the core would not fail before the surface.

It is important to note that the yield condition and irreversible flow in the SSDEM are consistent with the Drucker-Prager yield criterion and an associated flow rule, respectively \citep{Sanchez2012}, while the analytical mode considers the Mohr-Coulomb yield criterion and an associated flow rule. In the present study, we use these different yield criteria to show how they are consistent with each other in the present problem. 

\begin{figure}[hb]
  \centering
  \includegraphics[width=6in]{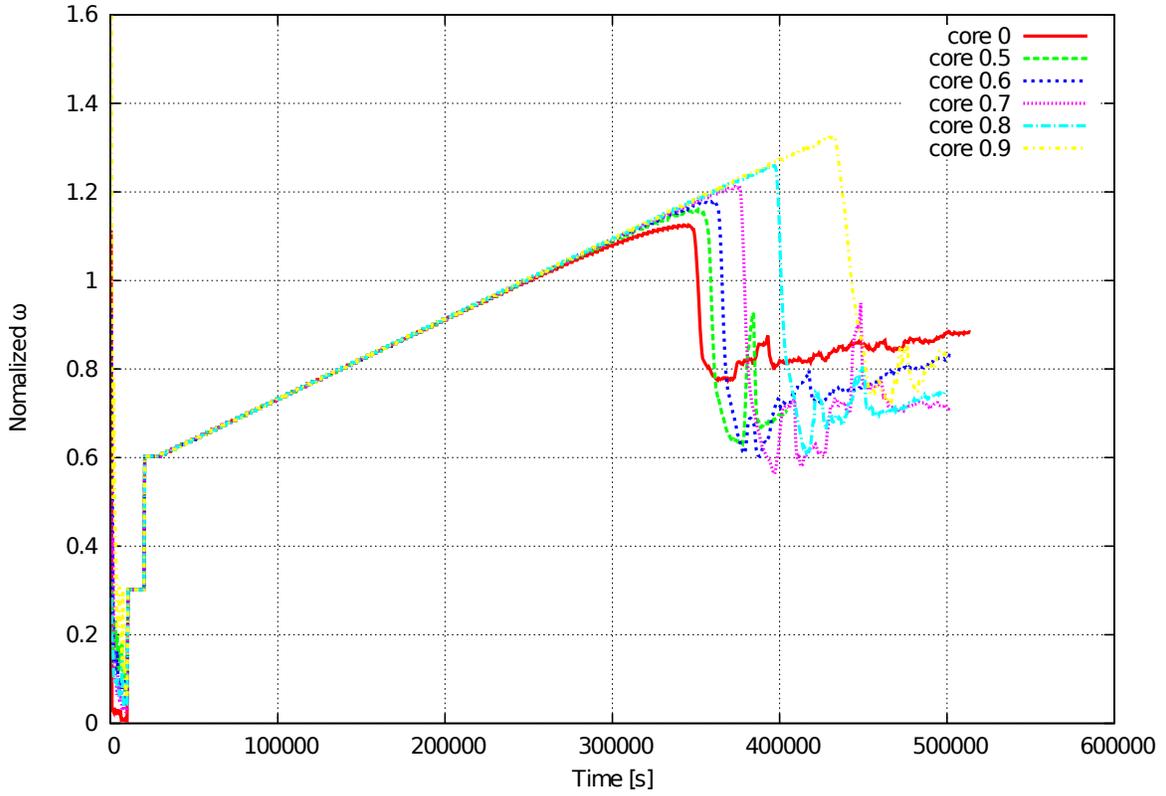}
  \caption{Time evolution of the angular velocity of the test body. The colors indicate different core radii; the red, the green, blue, magenta, light blue and yellow show normalized core radii of 0, 0.5, 0.6, 0.7, 0.8 and 0.9, respectively. The angular velocity of the test body is incremented in steps of a normalized spin rate of $5.5\times 10^{-3}$ every 3000 s.}
  \label{Fig:NormalizedSpin}
\end{figure}

\begin{figure}[hb]
  \centering
  \includegraphics[width=6in]{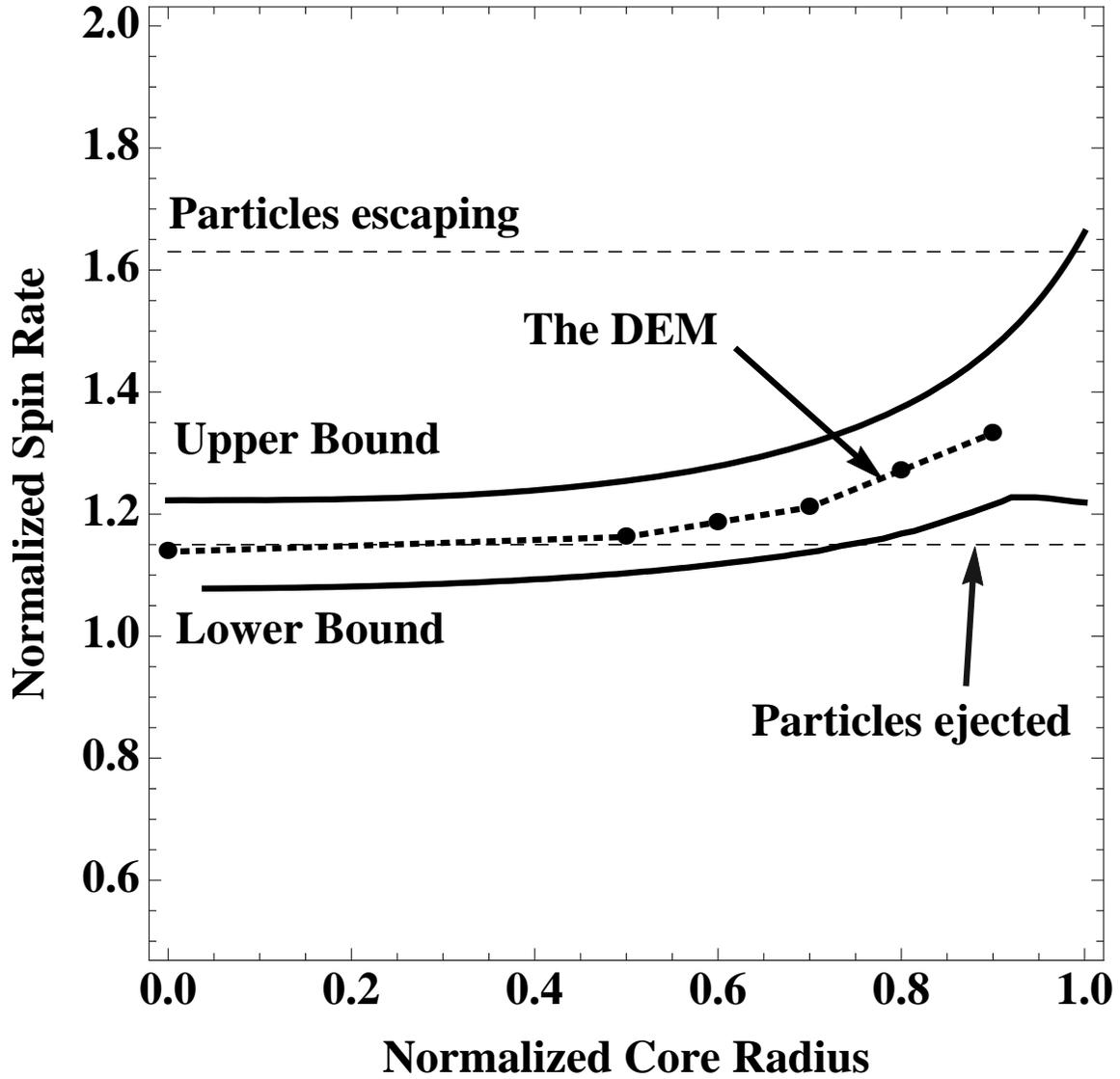}
  \caption{Comparison between the analytical model (the solid lines) and the SSDEM (the dotted line). On these curves, the cohesion is fixed at 0.10. The SSDEM result is perfectly enclosed by the upper and lower bound conditions derived by the analytical model. The lower dashed line indicates the condition at which particles are just shed from the surface, while the upper dashed line describes that when particles escape.}
  \label{Fig:Comparison}
\end{figure}

We first compare the failure conditions obtained from the SSDEM and the analytical models. Figure \ref{Fig:NormalizedSpin} shows the spin-up profile for the cases of normalized core radii of 0, 0.5, 0.6, 0.7, 0.8 and 0.9, respectively. The peak spin rates for these cases are 1.13, 1.16, 1.18, 1.21, 1.27 and 1.33, respectively. After experiencing their peak spin rates, the test body cannot reach this spin rate again. This comes from the fact that the shape has already changed irreversibly, causing the test body to be more susceptible to failure. An increase in spin rate induces further deformation with the consequent slow-down in spin rate to conserve angular momentum. Figure \ref{Fig:Comparison} shows the spin rate at a constant surface cohesion of 0.10 as a function of the core radius. The solid lines are obtained by the analytical model, while the dotted line is given by the SSDEM. The simulation results are located between the lower and upper bound conditions calculated by the analytical model, showing the consistency of these two approaches

Next, we observe the deformation of the test body that occurs right after the rapid spin-rate drop to investigate the initial process of its failure mode. Figure \ref{Fig:core00} shows the deformation mode of the test body with a normalized core radius of 0. Figure \ref{Fig:core00a} shows the 3-dimensional depiction of the aggregate after failure. The color shows the latitude of the test body. Figures \ref{Fig:core00b} through \ref{Fig:core00d} show the deformation vectors over the thin slices that include the center of mass and which are perpendicular to the $x$ through $z$ axes, respectively. Figures \ref{Fig:core00b} and \ref{Fig:core00c} (views from +x and +y) show how the poles of the aggregate push inwards towards the center, helping in the formation of an equatorial bulge.  Figure \ref{Fig:core00d} (view from +z) shows how the aggregate deforms outwards from the center.  This deformation mode corresponds to the deformation of 1950 DA \citep{Hirabayashi2015_DA} and could lead to its catastrophic disruption of the internal core. A possible disruption mode for this case will be discussed in Section \ref{Sec:Discussion}. 

\begin{figure}[ht!]
	\begin{center}
		\subfigure[]{
         		\label{Fig:core05a}	
		\includegraphics[width=2.85in]{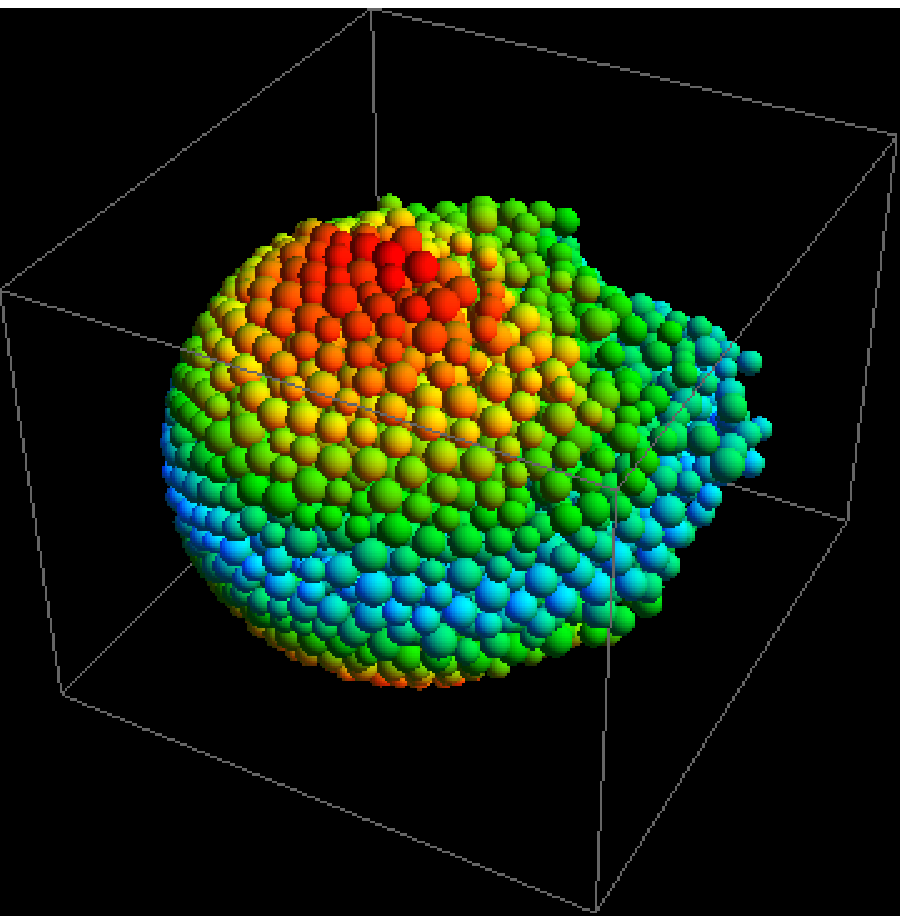}
          	}
		\subfigure[]{
         		\label{Fig:core05b}	
		\includegraphics[width=3in]{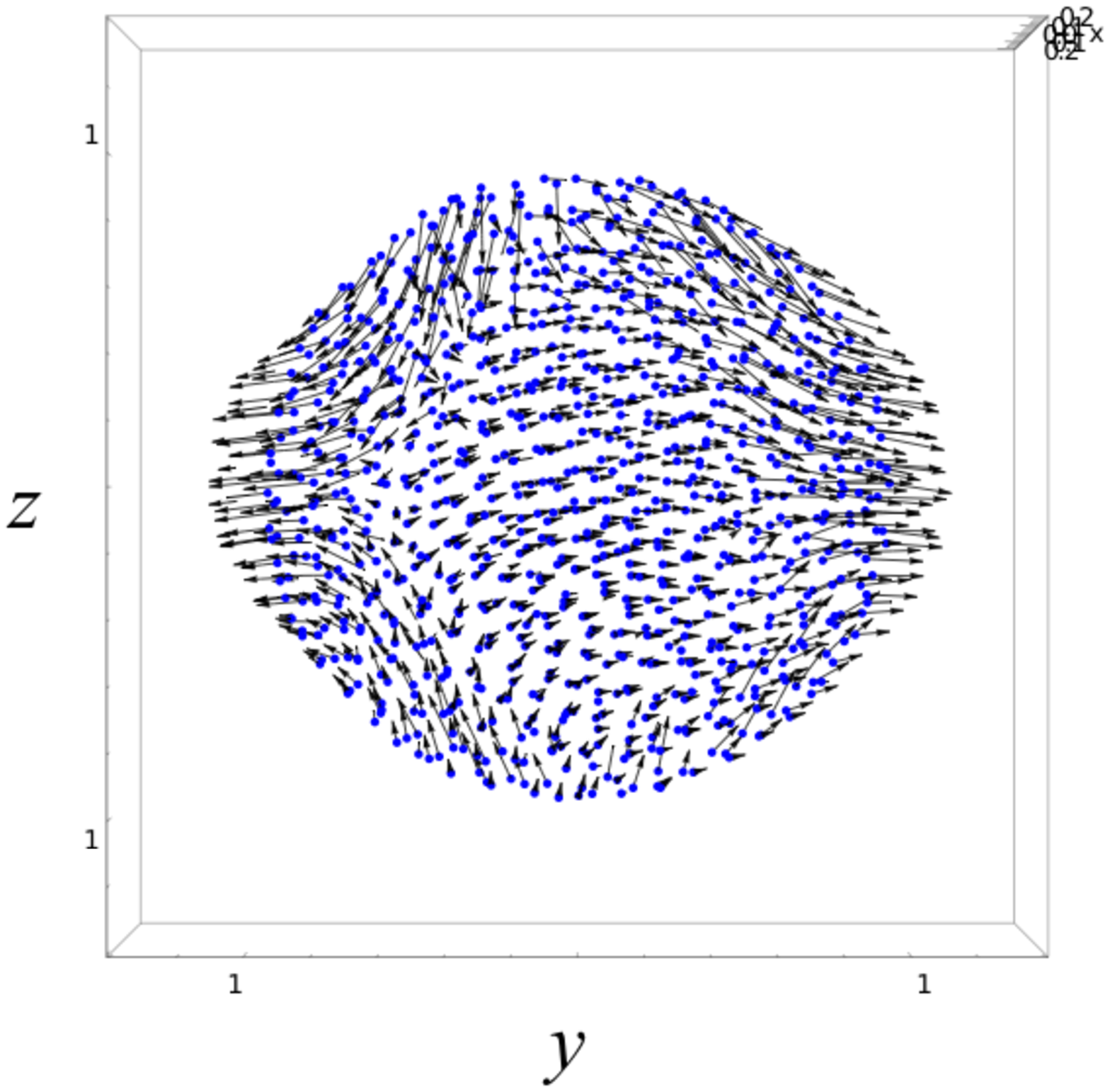}
          	} \\
		\subfigure[]{
         		\label{Fig:core05c}	
		\includegraphics[width=2.85in]{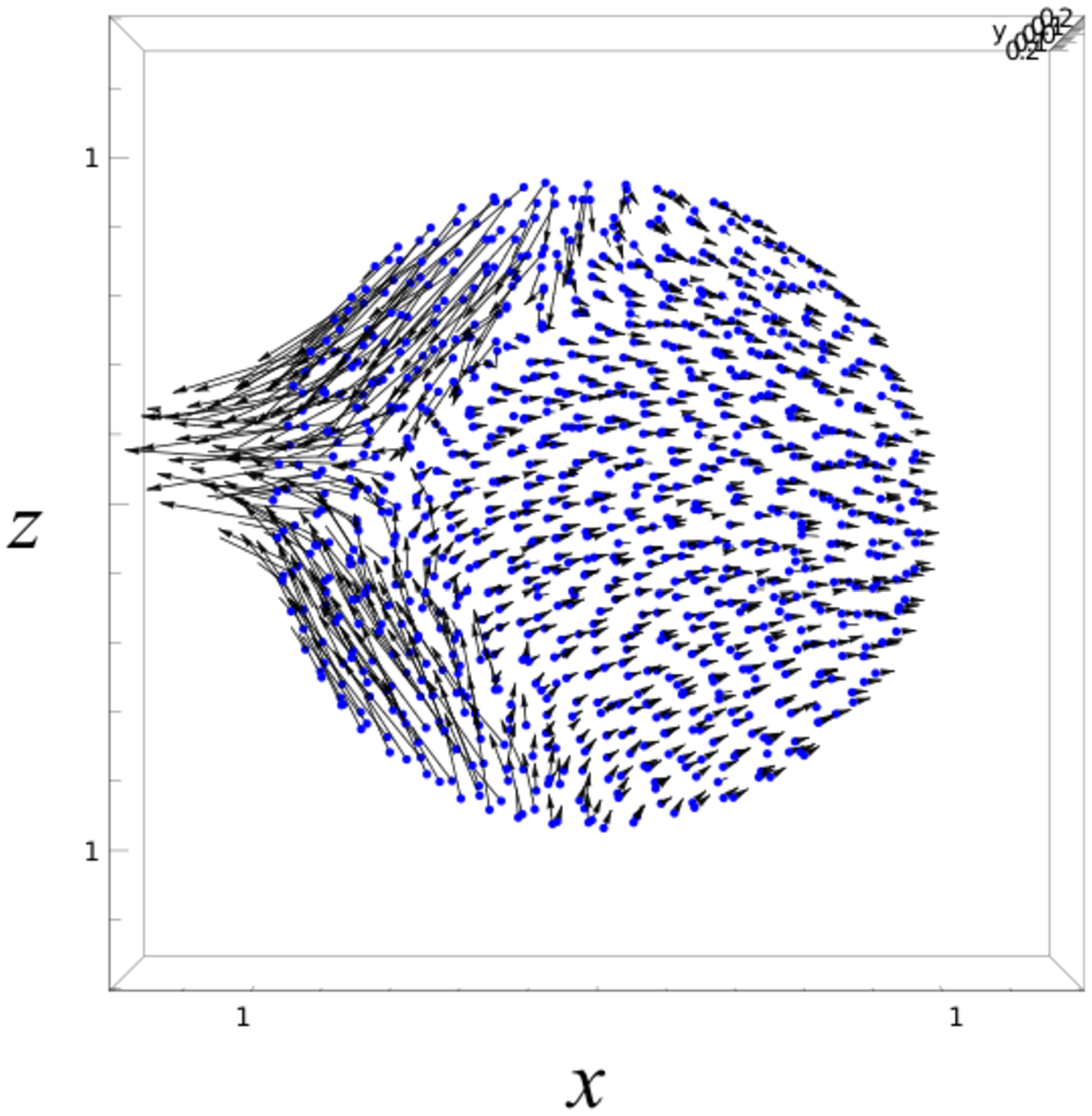}
          	}
		\subfigure[]{
         		\label{Fig:core05d}	
		\includegraphics[width=3in]{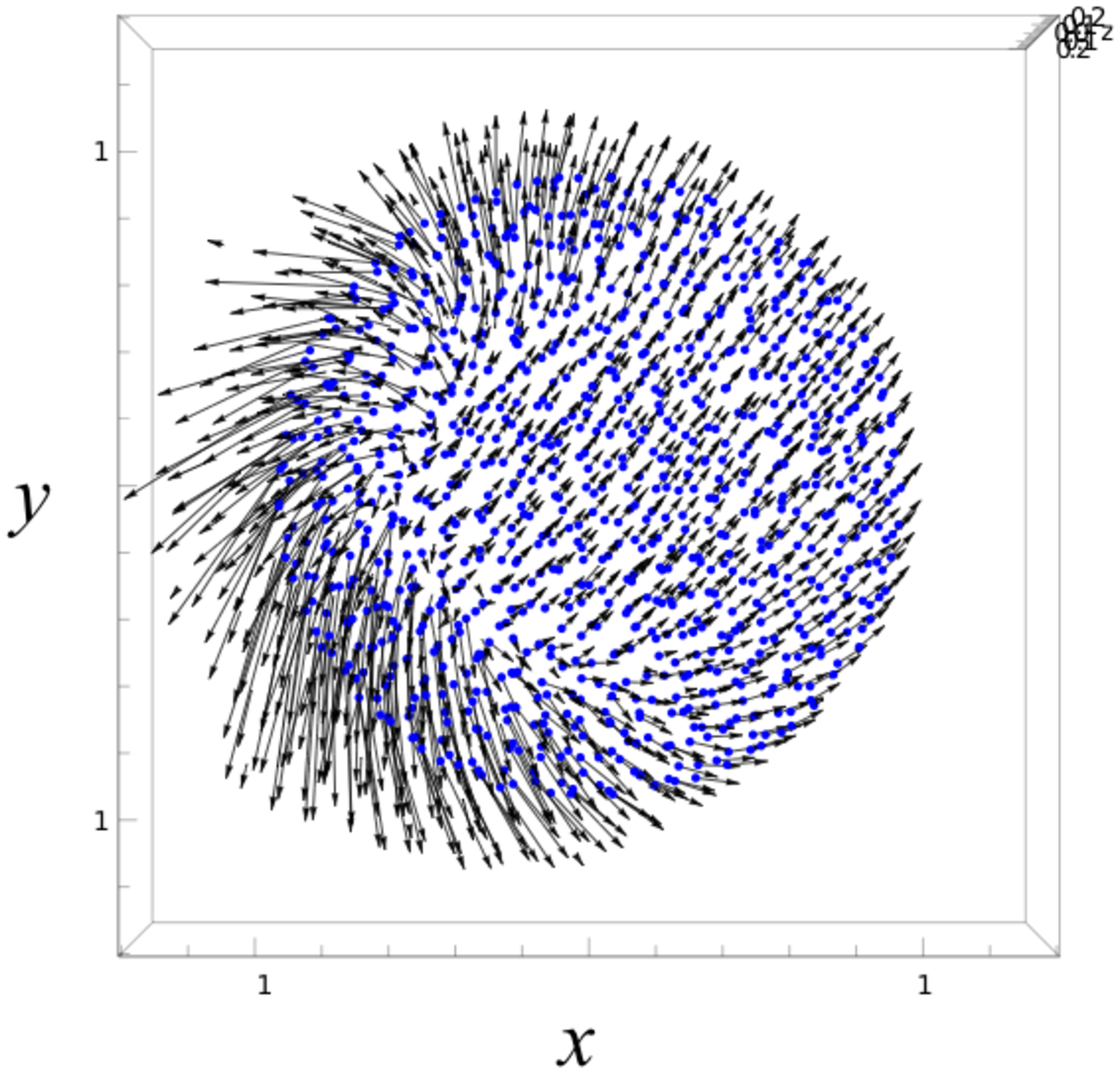}
          	} \\	
	\caption{Deformation of the test body with a normalized core radius of 0.5. The normalized cohesion of the surface shell and that of the internal core are 0.1 and 0.5, respectively. The normalized critical spin rate is 1.16. The definitions of the figures are the same as Figure \ref{Fig:core00}.}
	\label{Fig:core05}
	\end{center}
\end{figure}

\begin{figure}[ht!]
	\begin{center}
		\subfigure[]{
         		\label{Fig:core09a}	
		\includegraphics[width=2.85in]{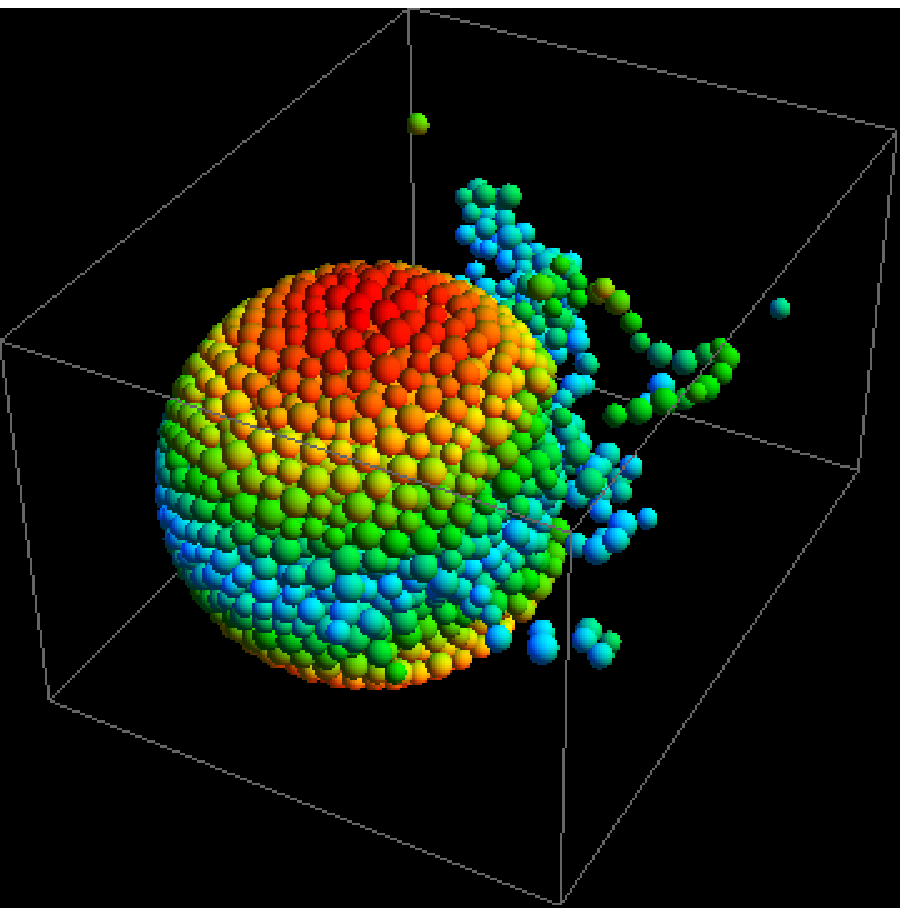}
          	}
		\subfigure[]{
         		\label{Fig:core09b}	
		\includegraphics[width=3in]{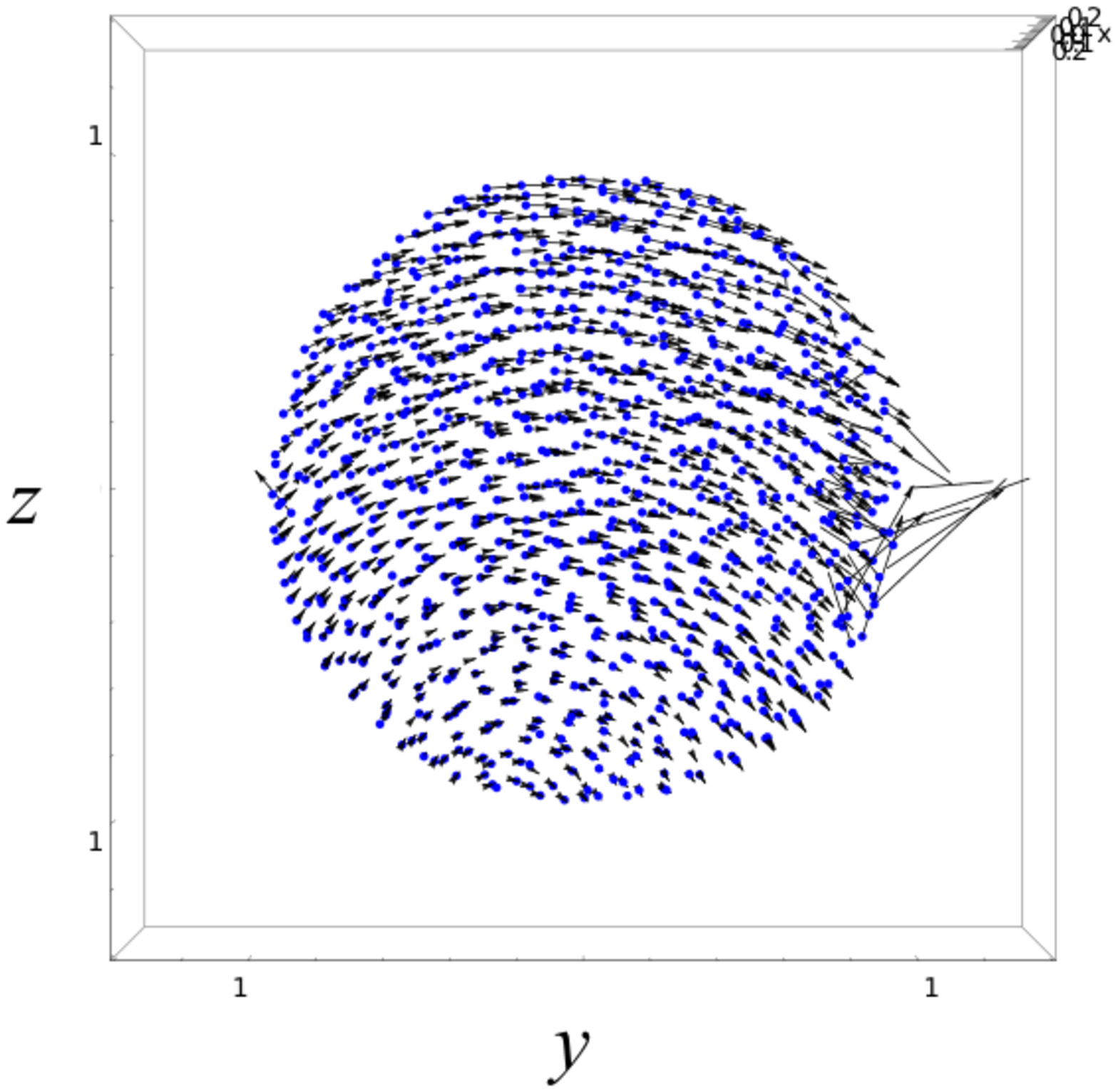}
          	} \\
		\subfigure[]{
         		\label{Fig:core09c}	
		\includegraphics[width=2.85in]{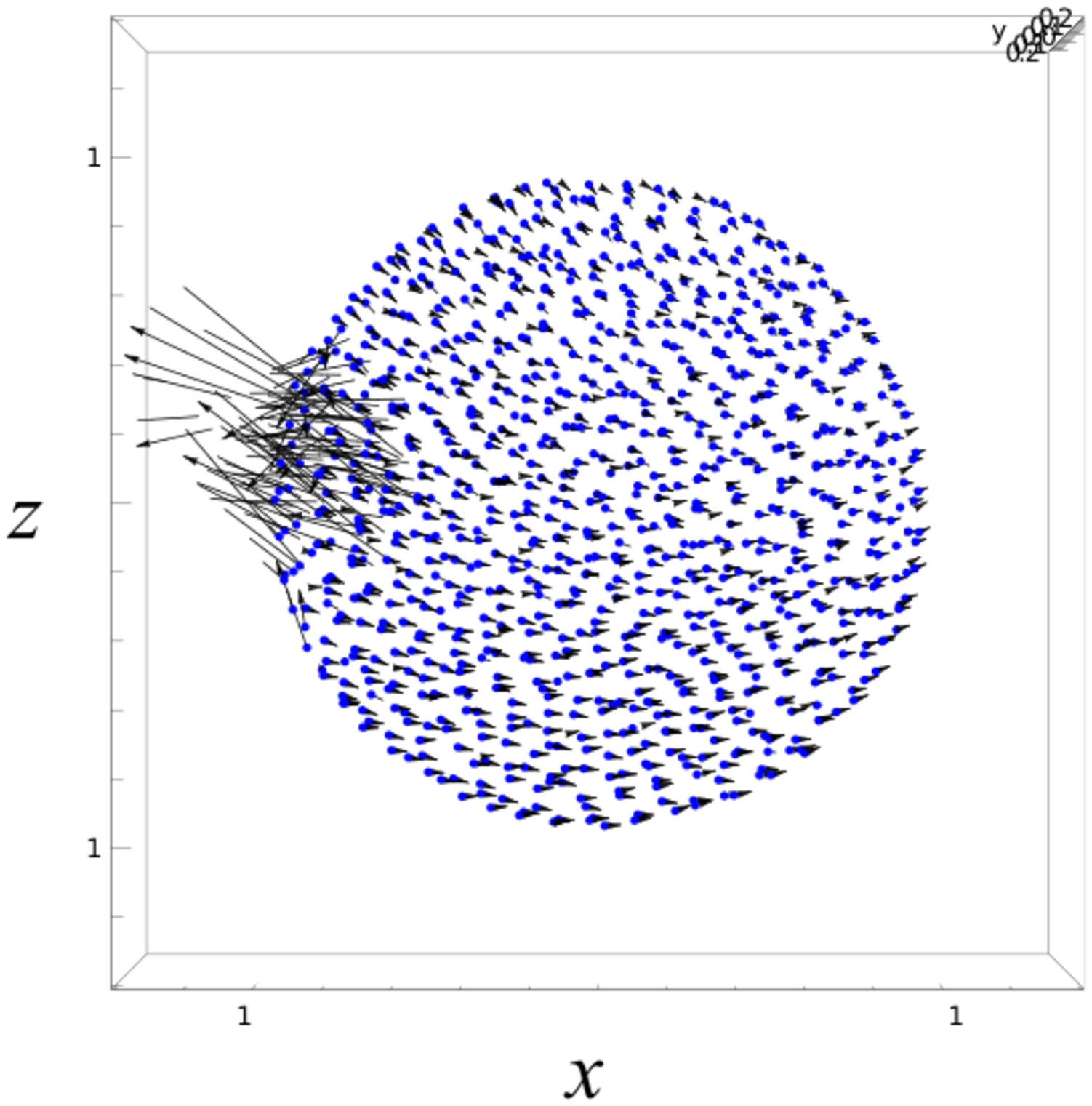}
          	}
		\subfigure[]{
         		\label{Fig:core09d}	
		\includegraphics[width=3in]{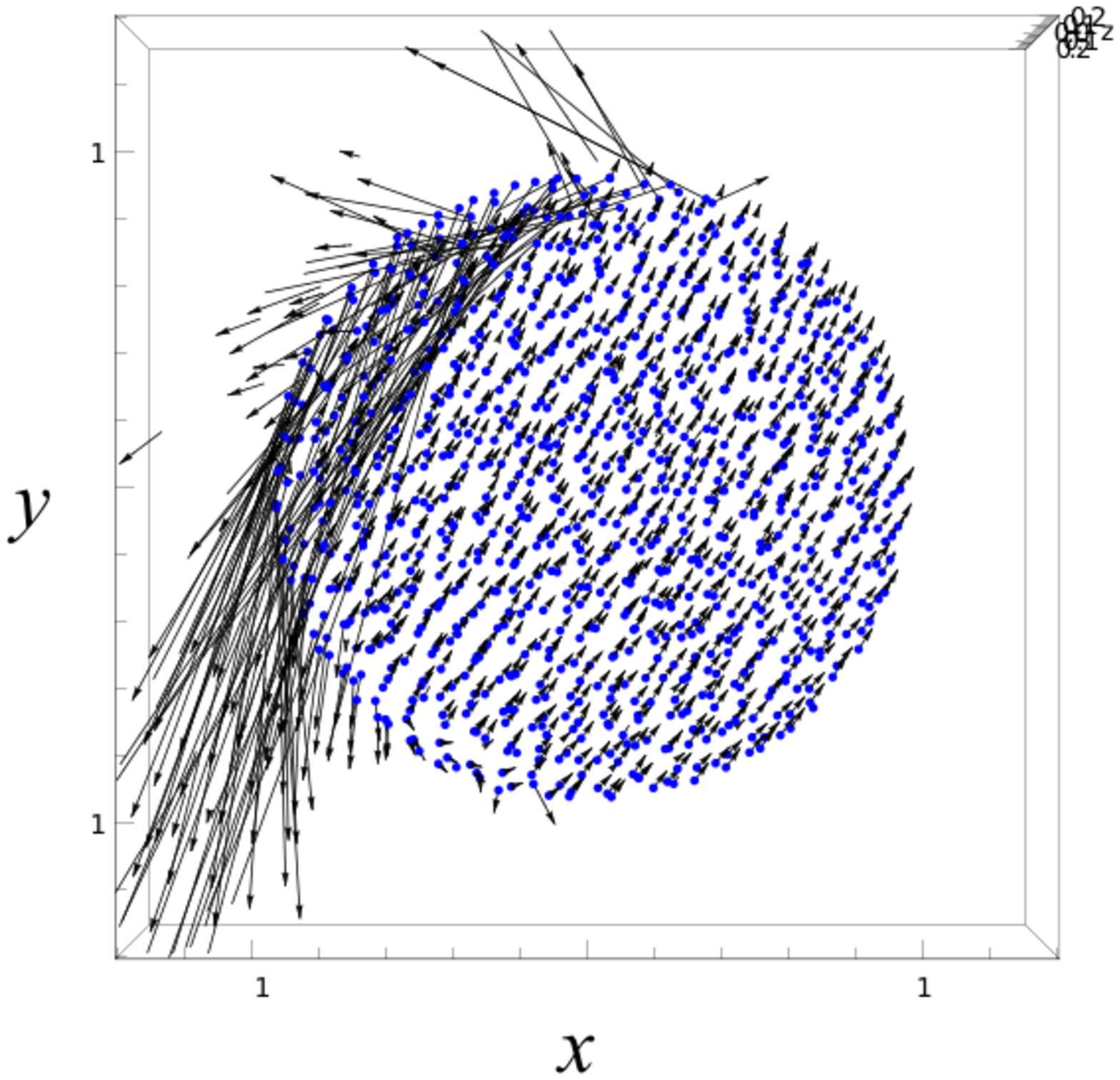}
          	} \\	
	\caption{Deformation of the test body with a normalized core radius of 0.9. The normalized cohesion of the surface shell and that of the internal core are 0.1 and 0.5, respectively. The normalized critical spin rate is 1.33. The definitions of the figures are the same as Figure \ref{Fig:core00}. }
	\label{Fig:core09}
	\end{center}
\end{figure}

As seen in Figures \ref{Fig:core05} and \ref{Fig:core09}, on the other hand, the effect of the internal core on the deformation is obvious. The deformation of the internal core is minimal compared to the homogeneous (zero-core) case, while the surface shell fails structurally\footnote{The maximum deformation of the core occurs in the equatorial plane. For a normalized core of 0.5, the deformation is up to 3.5\% in size, which is negligible compared to the surface deformation. As the core size becomes larger, the deformation becomes smaller.}. For the case of a normalized radius of 0.5 (Figure \ref{Fig:core05}), after the surface shell fails, the particles there flow over the surface to reach the equatorial region. It is found that the failure region is not global, but local. For the case of a normalized radius of 0.9 (Figure \ref{Fig:core09}), the failure region becomes smaller than that for the case of a normalized radius of 0.5 and mainly occurs near the equatorial region. 

Now we analyze Figures \ref{Fig:core00} through \ref{Fig:core09} more closely to understand the effect of the internal core. These cases show the following three characteristics. First, the flow of particles is asymmetric with respect to the rotation axis and is symmetric with respect to the the equatorial plane, causing a ridge-like shape. Second, the region with the greatest deformation is flat and is diametrically opposed to the region that is least deformed and that is still very round. Third, points on the equatorial region between these two extremes show intermediate degrees of deformation/stretching (oblateness). These characteristics become less and less pronounced as the core grows in size. For a radius of 0.9, there is no deformation towards an oblate shape and the particles of the equator are simply ejected from the main body. The deformation of the test body happens only due to the failure of the surface shell, so the thicker it is, the greater the deformation. This implies that for the case of a normalized core radius of 0, the deformation is global \citep{Sanchez2012}. 

Our simulations, in no case show a deformation towards a symmetric oblate body and instead favor a wedge-like shape. As homogeneous as we would like our aggregates to be, by construction they will present internal weak points where the body (as any real aggregate) can fail more easily.  This is the reason for local nature of the region of greater deformation.  Once the aggregate deforms in one region, its internal stress is diminished as the angular velocity has to decrease to conserve angular momentum.  In this paper we have only focused on the initial phase on deformation so that the dynamics of surface shedding can be understood. However, future efforts will be directed towards understanding of its stable shape.

The simulations also show the dynamics of surface shedding. In Figure \ref{Fig:Comparison}, the lower dashed line represents the condition at which the particles are just lofted from the surface, i.e., 1.15. If the spin rate is faster than this line and the surface shell already fails, surface shedding should occur. The upper dashed line represents the condition at which particles escape to infinity, i.e., 1.63. According to this figure, if the core radius is 0, particles cannot be shed. For the cases of core radii of 0.5 and larger, particles are lofted. However, since all the cases are below the upper dashed line, particles cannot reach escape velocity. They would either stay in orbit around the primary body or come back to its surface. Potentially, an absolute increase in the cohesion of the shell and interior would enable shed particles to immediately escape upon their being shed.

\section{Discussion}
\label{Sec:Discussion}
\begin{figure}[hb]
  \centering
  \includegraphics[width=6in]{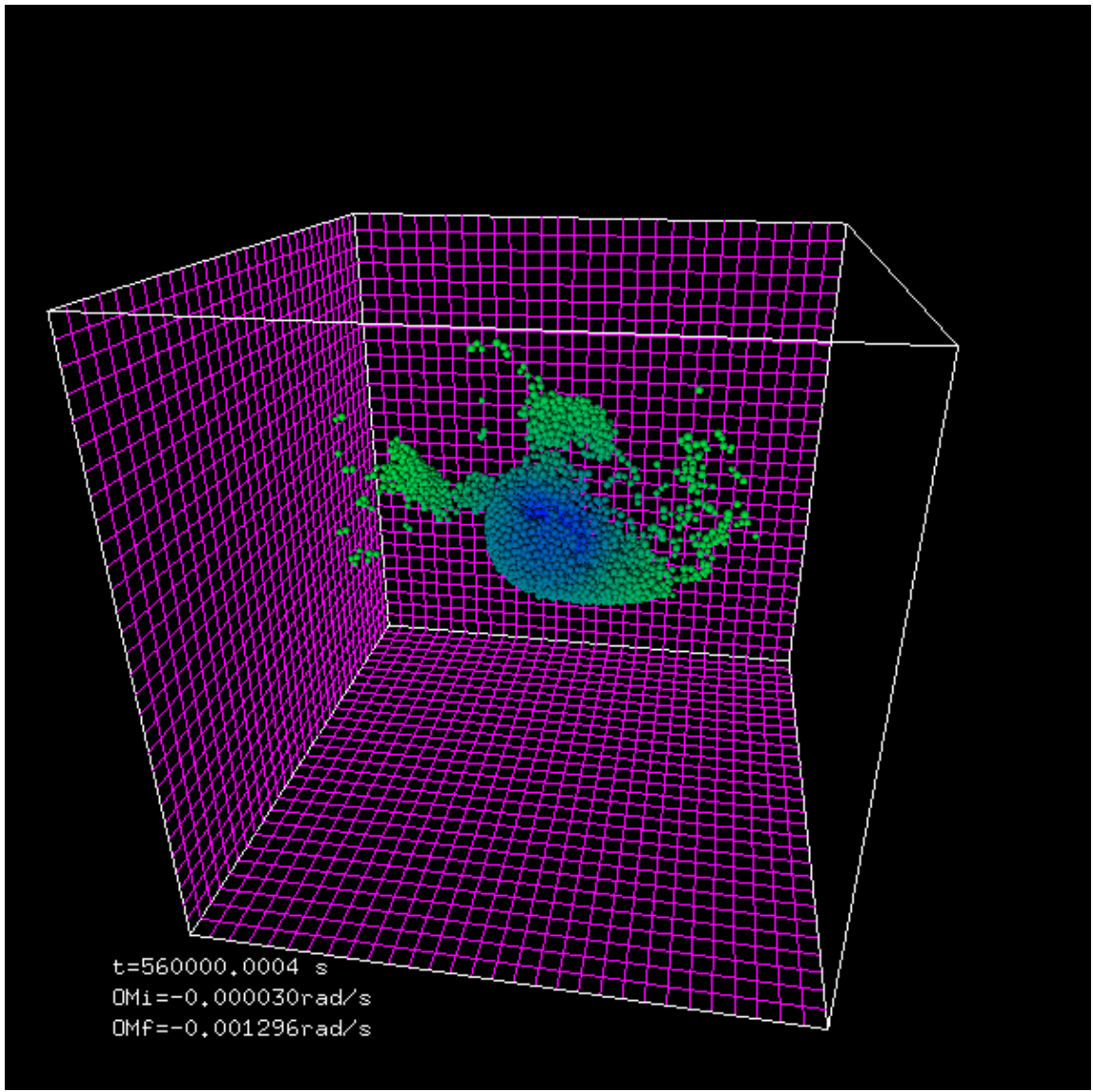}
  \caption{Example of the final failure mode of a zero-core spheroid \citep{sanchez-dps2014, sanchez-acm2014}. The normalized cohesion of the whole body is 0.5, and the normalized critical spin rate is 1.44. The simulation time is 560000s, which is 10000s longer than the time when the initial failure mode is observed ($\sim$460000s). The progenitor body is completely breaking into several large components.}
  \label{Fig:ZeroCore_Ultimate}
\end{figure}

In this paper, we have proposed a possible structure for a rubble pile asteroid that experiences surface shedding. It is important to consider what we know about failure modes from ground observations. \cite{Holsapple2010} found that if the internal structure is uniformly distributed, the original body completely deforms before surface shedding. This predicts that a homogeneous spheroid would never have surface shedding, but could experience a catastrophic breakup at high enough cohesion. \cite{sanchez-dps2014, sanchez-acm2014} simulated the zero core case for 560000s, which is 10000s longer than the initial failure time ($\sim 460000$s). The cohesion was fixed at 0.50, which is the same cohesion used for the internal core in the previous sections. Figure \ref{Fig:ZeroCore_Ultimate} shows a snapshot of the failure mode at 560000s. The test body catastrophically breaks up into multi-components. This mode is clearly distinguished from surface shedding. The surface shedding mode originates from the failure of the surface shell, causing granular flow.

Several asteroids recently observed are now considered to have undergone surface shedding. Asteroid P/2013 P5 (PANSTARRS) was reported to experience its episodic dust ejection \citep{Jewitt2013, Jewitt2015_P5}. This asteroid is orbiting near the inner edge of the asteroid belt in the vicinity of the Flora family of $S$-type asteroids. It implies that sublimation of water ice is unlikely to occur on this asteroid \citep{Jewitt2013}. The size of dust particles was up to a few millimeters, negligibly smaller than the nucleus, and the ejection velocity was less than 1 m/s \citep{Jewitt2015_P5}. From these observational results, the episodic event of this asteroid is currently considered to come from surface shedding. Asteroids 133P/Elst-Pizarro and (62412) 2000 SY178 have relatively rapid spin periods of 3.47 hr and 3.33 hr, respectively, and had experienced mass ejection. Although the mass ejection might have been driven by the seasonal activity, i.e., sublimation of water ice, rotational instability is also considered to help their activity \citep{Hsieh2004, Hsieh2010, Sheppard2015}. Some remarkable observations related to their activity include: (1) the size of ejecta is small, (2) the mass ejection is negligible compared to the nucleus and (3) the ejection velocity is relatively small. These facts contradict the catastrophic breakup of the coreless spheroid observed in Figure \ref{Fig:ZeroCore_Ultimate}, but is consistent with our model of an internal core. It implies that these active asteroids may have internal cores which prevent them from breaking up catastrophically. 

Additionally, simulations allowed us to observe that the failure due to rotational deformation has a local nature. The local deformation of the surface shell is dependent on its structure, especially on the spatial distribution of particles which is translated in the spatial variation of porosity and the possible presence of fractures. Even though the settling procedure, devoided of cohesive and frictional forces, encourages structural homogeneity, this is not perfect and, at high enough spin rates, granular flows always occur locally.  We consider this to be the initial deformation mode.  Though long-term deformation modes are not the focus of this paper, there are two possible scenarios for them.  In the first scenario, material flows occur globally, causing the formation of a uniform equatorial ridge. This scenario corresponds to the result by obtained \cite{Walsh2008,Walsh2012}. In the second scenario, a local deformation leads to further structural bias, causing the further emergence of local material flows. We will conduct further investigations for this problem in the future.

The formation process of a strong core is beyond our scope; thus, we do not quantitatively analyze it in this paper. However, there may be two likely scenarios. The first scenario is that at an accretion process after a catastrophic impact, due to the difference of energy dissipation, large boulders and rocks could accumulate first and then small particles might cover these large objects. Such a formation process could produce bodies with layers of different strength similar to the model we have here analyzed (Durda and Walsh, 2015 personal communication). Note that since mechanical strength depends on many different parameters such as bulk density \citep{Hirabayashi2014B}, friction angle and cohesion, we cannot conclude that such a structure gives a strong core. We leave this problem for future works. The second scenario is based on the assumption that the distribution of loose materials is uniform initially. This condition leads to compression of loose particles in the interior. When such particles irreversibly move out of their original locations, the volumetric strain decreases as compression increases \citep{Lambe1969,Nedderman1992}. Because of the self-gravity, the interior experiences larger compression than the surface. As a result, particles in the interior are highly packed. Since higher compaction gives higher cohesion \citep{Lambe1969}, this scenario could produce a strong core in an asteroid. 

Finally, we compare the HSDEM by \cite{Walsh2008, Walsh2012} with our model. \cite{Walsh2008, Walsh2012} observed surface shedding from a rubble pile body with a homogeneous hexagonal closed packing (HCP) structure\footnote{They defined several initial packing configurations to control the friction angle.}, while we concluded that a body should have an internal core to have surface shedding. This different result comes from the fact that the simulations carried out by \cite{Walsh2008,Walsh2012} and our model may have different deformation features. If particles are arranged to form a crystalline lattice, each particle dislocation is highly controlled by its closest neighbors and the deformation is not isotropic \citep{Ashby2014}. However, even for such a configuration, particles sitting on the surface have open spaces where they can move freely; thus, their motion is less restricted than if they were in the interior. Consequently, the initial HCP configuration could cause anisotropic flows and provide their aggregates with a strong core and a weak surface shell implicitly. Note that \cite{Walsh2012} also investigated a near-fluid case that was modeled by the bi-modal size distribution (a randomly packed aggregate) and obtained a pancake shape, which might indicate an isotropic deformation mode.

\section{Conclusion}
This paper quantitatively analyzed the effect of the internal core of a rubble pile spheroid on the mechanism of surface shedding. On the assumption that the friction angle and the bulk density are uniform over the whole volume, we introduced a simple surface shedding model. It is composed of two different cohesion layers: the internal core and the surface shell. The analytical method was constructed by the lower and upper bound theorems in limit analysis, while the numerical method was based on the SSDEM by S\'anchez and Scheeres. The failure conditions derived by the analytical model were consistent with those by the SSDEM. The results showed that an aggregate with homogeneous cohesion would always fails internally before it fails superficially. To have surface shedding, therefore, the progenitor body needed a strong core so that only the surface region failed structurally. The SSDEM also exposed the dynamics of the initial failure phase. The primary result was that for all the cases, a rubble pile spheroid would fail locally through a severe material flow on its surface. This came from the fact that since the initial body configuration was not perfectly uniform, which we believe is more realistic, some surface regions are susceptible to fail earlier than others. Further research would be able to provide a better understanding of the formation of unique shapes of asteroids and the mechanisms that originate the activity observed in active asteroids.

\acknowledgments
The authors wish to thank Dr. Kevin Walsh and Dr. Dan Durda at SwRI Boulder and Dr. Dave Jewitt at UCLA for their useful comments on the present work. 

\appendix

\section{Averaged Stresses over the Core and the Surface Shell}
\label{App:AvgStress}
The averaged stresses over the internal core and the surface shell are given by the integrations of Equations (\ref{Eq:sigxx}) through (\ref{Eq:sigyz}) over these volumes. Since the shear stress components given by Equations (\ref{Eq:sigxy}) through (\ref{Eq:sigyz}) are trigonometric functions, they become zero by the integration. On the other hand, the normal stress components, Equations (\ref{Eq:sigxx}) through (\ref{Eq:sigzz}), have non-zero values. Using the indices notations defined above, we obtain the averaged normal stresses for the surface shell as
\begin{eqnarray}
\bar \sigma_{1s} &=& \frac{3}{4 \pi (R^3 - R_b^3)} \int_{R_b}^1 \int_{- \frac{\pi}{2}}^{\frac{\pi}{2}} \int_{0}^{2 \pi} \sigma_{1} (r, \theta, \psi) r^2 \cos \theta d\psi d\theta dr, \nonumber \\
&=& \frac{3 k_1}{1 - R_b^3} \left(\frac{2}{15} - \frac{R_b^3}{3} + \frac{R_b^5}{5} \right) + \frac{b_x - 2 k_1}{5} \frac{1 - R_b^5}{1 - R_b^3}, \\
\bar \sigma_{3s} &=& \frac{3}{4 \pi (R^3 - R_b^3)} \int_{R_b}^1 \int_{- \frac{\pi}{2}}^{\frac{\pi}{2}} \int_{0}^{2 \pi} \sigma_{3} (r, \theta, \psi) r^2 \cos \theta d\psi d\theta dr, \nonumber \\
&=& \frac{3 k_9}{1 - R_b^3} \left(\frac{2}{15} - \frac{R_b^3}{3} + \frac{R_b^5}{5} \right) + \frac{b_z - 2 k_9}{5} \frac{1 - R_b^5}{1 - R_b^3},
\end{eqnarray}
and those for the internal core as
\begin{eqnarray}
\bar \sigma_{1c} &=& \frac{3}{4 \pi R_b^3} \int_0^{R_b} \int_{- \frac{\pi}{2}}^{\frac{\pi}{2}} \int_{0}^{2 \pi} \sigma_{11} (r, \theta, \psi) r^2 \cos \theta d\psi d\theta dr, \nonumber \\
&=& k_1 \left(1 - R_b^2 \right) + \frac{b_x R_b^2}{5}, \\
\bar \sigma_{3c} &=& \frac{3}{4 \pi R_b^3} \int_0^{R_b} \int_{- \frac{\pi}{2}}^{\frac{\pi}{2}} \int_{0}^{2 \pi} \sigma_{11} (r, \theta, \psi) r^2 \cos \theta d\psi d\theta dr, \nonumber \\
&=& k_9 \left(1 - R_b^2 \right) + \frac{b_z R_b^2}{5}. 
\end{eqnarray}
Note that $\bar \sigma_{1s} = \bar \sigma_{2s}$ and $\bar \sigma_{1c} = \bar \sigma_{2c}$.  Since the shear stress components are zero, these components are considered to be the principal stress components.

\bibliographystyle{apj}

\begin{thebibliography}{47}
\expandafter\ifx\csname natexlab\endcsname\relax\def\natexlab#1{#1}\fi

\bibitem[{Ai {et~al.}(2011)Ai, Chen, Rotter, \& Ooi}]{ai-chen2011}
Ai, J., Chen, J.-F., Rotter, J.~M., \& Ooi, J.~Y. 2011, Powder Technology, 206,
  269

\bibitem[{Ashby {et~al.}(2014)Ashby, Shercliff, \& Cebon}]{Ashby2014}
Ashby, M.~F., Shercliff, H., \& Cebon, D. 2014, Materials: Engineering,
  Science, Processing and Design, 3rd edn. (Elsevier)

\bibitem[{Chen \& Han(1988)}]{Chen1988}
Chen, W.~F. \& Han, D.~J. 1988, Plasticity for Structural Engineers
  (Springer-Verlag)

\bibitem[{Chen \& Liu(1990)}]{Chen1990}
Chen, W.-F. \& Liu, X.~L. 1990, Limit analysis in soil mechanics (Elsevier
  Science Publishers B. V.)

\bibitem[{Cundall(1971)}]{cundall1971}
Cundall, P. 1971, in Proceedings of the International Symposium on Rock
  Mechanics

\bibitem[{Cundall \& Hart(1992)}]{cundall1992}
Cundall, P.~A. \& Hart, R.~D. 1992, Engineering Computations, 9, 101

\bibitem[{Dobrovolskis(1982)}]{Dobrovolskis1982}
Dobrovolskis, A.~R. 1982, Icarus, 52, 136

\bibitem[{{\v D}urech {et~al.}(2008){\v D}urech, Vokrouhlick{\'y}, Kaasalainen,
  Higgins, Krugly, Gaftonyuk, Shevchenko, Chiorny, Hamanowa, Reddy, \&
  Dyvig}]{Durech2008}
{\v D}urech, J., Vokrouhlick{\'y}, D., Kaasalainen, M., Higgins, D., Krugly,
  Y.~N., Gaftonyuk, N.~M., Shevchenko, V.~G., Chiorny, V.~G., Hamanowa, H.,
  Reddy, V., \& Dyvig, R.~R. 2008, A\&A, 489, L25

\bibitem[{Guibout \& Scheeres(2003)}]{Guibout2003}
Guibout, V. \& Scheeres, D. 2003, Celestial Mechanics and Dynamical Asronomy,
  87, 263

\bibitem[{Hainaut \& Snodgrass(2014)}]{Hainaut2014ACM}
Hainaut, O. \& Snodgrass, C. 2014, in Asteroids, Comets, Meteors

\bibitem[{Hainaut {et~al.}(2014)Hainaut, Boehnhardt, Snodgrass, Meech, Deller,
  Gillon, Jehin, Kuehrt, Lowry, Manfroid, Micheli, Mottola, Opitom, Vincent, \&
  Wainscoat}]{Hainaut2014}
Hainaut, O.~R., Boehnhardt, H., Snodgrass, C., Meech, K.~J., Deller, J.,
  Gillon, M., Jehin, E., Kuehrt, E., Lowry, S.~C., Manfroid, J., Micheli, M.,
  Mottola, S., Opitom, C., Vincent, J.-B., \& Wainscoat, R. 2014, A\&A, 563,
  A75

\bibitem[{Harris {et~al.}(2009)Harris, Fahnestock, \& Pravec}]{Harris2009}
Harris, A.~W., Fahnestock, E.~G., \& Pravec, P. 2009, Icarus, 199, 310

\bibitem[{Herrmann \& Luding(1998)}]{herr1}
Herrmann, H. \& Luding, S. 1998, Continuum Mechanics and Thermodynamics, 10,
  189, 10.1007/s001610050089

\bibitem[{Hirabayashi(2014)}]{Hirabayashi2014B}
Hirabayashi, M. 2014, Icarus, 236, 178

\bibitem[{Hirabayashi \& Scheeres(2014)}]{Hirabayashi2014}
Hirabayashi, M. \& Scheeres, D. 2014, The astrophysical journal, 780

\bibitem[{Hirabayashi \& Scheeres(2015)}]{Hirabayashi2015_DA}
---. 2015, The Astrophysical Journal Letters, 798, L8

\bibitem[{Holsapple(2001)}]{Holsapple2001}
Holsapple, K.~A. 2001, Icarus, 154, 432

\bibitem[{Holsapple(2004)}]{Holsapple2004}
---. 2004, Icarus, 172, 272

\bibitem[{Holsapple(2007)}]{Holsapple2007}
---. 2007, Icarus, 187, 500

\bibitem[{Holsapple(2008)}]{Holsapple2008A}
---. 2008, International journal of Non-Linear Mechanics, 43, 733

\bibitem[{Holsapple(2010)}]{Holsapple2010}
---. 2010, Icarus, 205, 430

\bibitem[{Hsieh {et~al.}(2004)Hsieh, Jewitt, \& Fern‡ndez}]{Hsieh2004}
Hsieh, H.~H., Jewitt, D., \& Fern‡ndez, Y.~R. 2004, The Astronomical Journal,
  127, 2997

\bibitem[{Hsieh {et~al.}(2010)Hsieh, Jewitt, Lacerda, Lowry, \&
  Snodgrass}]{Hsieh2010}
Hsieh, H.~H., Jewitt, D., Lacerda, P., Lowry, S.~C., \& Snodgrass, C. 2010,
  Monthly Notices of the Royal Astronomical Society, 403, 363

\bibitem[{Jewitt {et~al.}(2014)Jewitt, Agarwal, Li, Weaver, Mutchler, \&
  Larson}]{Jewitt2014B}
Jewitt, D., Agarwal, J., Li, J., Weaver, H., Mutchler, M., \& Larson, S. 2014,
  The astrophysical journal letters, 784, 1

\bibitem[{Jewitt {et~al.}(2013)Jewitt, Agarwal, Weaver, Mutchler, \&
  Larson}]{Jewitt2013}
Jewitt, D., Agarwal, J., Weaver, H., Mutchler, M., \& Larson, S. 2013, The
  astrophysical journal letters, 778, L21

\bibitem[{Jewitt {et~al.}(2015)Jewitt, Agarwal, Weaver, Mutchler, \&
  Larson}]{Jewitt2015_P5}
---. 2015, The Astrophysical Journal, 798, 109

\bibitem[{Jewitt {et~al.}(2010)Jewitt, Weaver, Agarwal, Mutchler, \&
  Drahus}]{Jewitt2010}
Jewitt, D., Weaver, H., Agarwal, J., Mutchler, M., \& Drahus, M. 2010, Nature,
  467, 817

\bibitem[{Kaasalainen {et~al.}(2007)Kaasalainen, Durech, Warner, Krugly, \&
  Gaftonyuk}]{Kaasalainen2007}
Kaasalainen, M., Durech, J., Warner, B.~D., Krugly, Y.~N., \& Gaftonyuk, N.~M.
  2007, Nature, 446, 420

\bibitem[{Lambe \& Whitman(1969)}]{Lambe1969}
Lambe, T.~W. \& Whitman, R.~V. 1969, Soil Mechanis (John Wiley \& Sons)

\bibitem[{Lowry {et~al.}(2007)Lowry, Fitzsimmons, Pravec, Vokrouhlick?,
  Boehnhardt, Taylor, Margot, Gal‡d, Irwin, Irwin, \& Kusnir‡k}]{Lowry2007}
Lowry, S.~C., Fitzsimmons, A., Pravec, P., Vokrouhlick?, D., Boehnhardt, H.,
  Taylor, P.~A., Margot, J.-L., Gal‡d, A., Irwin, M., Irwin, J., \& Kusnir‡k,
  P. 2007, Science, 316, 272

\bibitem[{Minton(2008)}]{Minton2008}
Minton, D.~A. 2008, Icarus, 195, 698

\bibitem[{Mitchell {et~al.}(1972)Mitchell, Houston, Scott, Costes, {Carrier, W.
  D., III}, \& Bromwell}]{Mitchell1972}
Mitchell, J.~K., Houston, W.~N., Scott, R.~F., Costes, N.~C., {Carrier, W. D.,
  III}, \& Bromwell, L.~G. 1972, in Proceedings of the Lunar Science Conference

\bibitem[{Nedderman(1992)}]{Nedderman1992}
Nedderman, R.~M. 1992, Statics and Kinematics of Granular Materials (University
  of Cambridge Press)

\bibitem[{Pravec {et~al.}(2007)Pravec, Harris, \& Warner}]{Pravec2007}
Pravec, P., Harris, A.~W., \& Warner, B.~D. 2007, Near Earth Objects, our
  Celestial Neighbors: Opportunity and Risk, Proceedings IAU Symposium, 167

\bibitem[{Rubincam(2000)}]{Rubincam2000}
Rubincam, D.~P. 2000, Icarus, 148, 2

\bibitem[{S\'anchez \& Scheeres(2011)}]{sanchez2011}
S\'anchez, P. \& Scheeres, D. 2011, The Astrophysical Journal, 727, 120

\bibitem[{S\'anchez \& Scheeres(2012)}]{Sanchez2012}
---. 2012, Icarus, 218, 876

\bibitem[{{S\'anchez} \& {Scheeres}(2014{\natexlab{a}})}]{sanchez-dps2014}
{S\'anchez}, P. \& {Scheeres}, D. 2014{\natexlab{a}}, in AAS/Division for
  Planetary Sciences Meeting Abstracts, Vol.~46, AAS/Division for Planetary
  Sciences Meeting Abstracts, 400.06

\bibitem[{{S\'anchez} \& {Scheeres}(2014{\natexlab{b}})}]{sanchez-acm2014}
{S\'anchez}, P. \& {Scheeres}, D. 2014{\natexlab{b}}, in Asteroids, Comets,
  Meteors 2014, ed. K.~{Muinonen}, A.~{Penttil{\"a}}, M.~{Granvik},
  A.~{Virkki}, G.~{Fedorets}, O.~{Wilkman}, \& T.~{Kohout}, 463

\bibitem[{S\'anchez \& Scheeres(2014)}]{sanchez2014}
---. 2014, Meteoritics \& Planetary Science, 49, 788

\bibitem[{Scheeres(2015)}]{Scheeres2015}
Scheeres, D. 2015, Icarus, 247, 1

\bibitem[{Sheppard \& Trujillo(2015)}]{Sheppard2015}
Sheppard, S.~S. \& Trujillo, C. 2015, The Astronomical Journal, 149, 44

\bibitem[{Silbert {et~al.}(2001)Silbert, Erta\ifmmode~\mbox{\c{s}}\else
  \c{s}\fi{}, Grest, Halsey, Levine, \& Plimpton}]{silbert}
Silbert, L.~E., Erta\ifmmode~\mbox{\c{s}}\else \c{s}\fi{}, D., Grest, G.~S.,
  Halsey, T.~C., Levine, D., \& Plimpton, S.~J. 2001, Phys. Rev. E, 64, 051302

\bibitem[{Snodgrass {et~al.}(2010)Snodgrass, Tubiana, Vincent, Sierks, Hviid,
  Moissl, Boehnhardt, Barbieri, Koschny, Lamy, Rickman, Rodrigo, Carry, Lowry,
  Laird, Weissman, Fitzsimmons, Marchi, \& the {OSIRIS}~team}]{Snodgrass2010}
Snodgrass, C., Tubiana, C., Vincent, J.-B., Sierks, H., Hviid, S., Moissl, R.,
  Boehnhardt, H., Barbieri, C., Koschny, D., Lamy, P., Rickman, H., Rodrigo,
  R., Carry, B., Lowry, S.~C., Laird, R. J.~M., Weissman, P.~R., Fitzsimmons,
  A., Marchi, S., \& the {OSIRIS}~team. 2010, Nature, 467, 814

\bibitem[{Taylor {et~al.}(2007)Taylor, Margot, Vokrouhlick\'y, Scheeres,
  Pravec, Lowry, Fitzsimmons, Nolan, Ostro, Benner, Giorgini, \&
  Magri}]{Taylor2007}
Taylor, P.~A., Margot, J.-L., Vokrouhlick\'y, D., Scheeres, D., Pravec, P.,
  Lowry, S.~C., Fitzsimmons, A., Nolan, M.~C., Ostro, S.~J., Benner, L. A.~M.,
  Giorgini, J.~D., \& Magri, C. 2007, Science, 316, 274

\bibitem[{Walsh {et~al.}(2008)Walsh, Richardson, \& Michel}]{Walsh2008}
Walsh, K.~J., Richardson, D.~C., \& Michel, P. 2008, Nature, 454, 188

\bibitem[{Walsh {et~al.}(2012)Walsh, Richardson, \& Michel}]{Walsh2012}
---. 2012, Icarus, 220, 514

\end{thebibliography}


\clearpage

\clearpage

\clearpage

\clearpage

\clearpage

\clearpage

\end{document}